\documentclass[12pt]{iopart}
\usepackage{graphicx}
\usepackage{cite}

\usepackage{soul}
\usepackage{xcolor}

\newcommand{\cc}{$^\circ\ $}

\newcommand{\ccnd}{$^\circ$}

\newcommand{\blockcomment}[1]{}

\begin{document}

\title{Collimated $ \gamma $-flash emission along the target surface irradiated by a laser at non-grazing incidence}

\author{M. Matys$^{1}$, P. Hadjisolomou$^{1}$, R. Shaisultanov$^{1}$, P. Valenta$^{1}$, M. Lama\v{c} $^{1}$, T. M. Jeong$^{1}$, J. P. Thistlewood$^{1,2}$, C. P. Ridgers$^{3}$, A. S. Pirozhkov$^{4}$  and  S.~V.~Bulanov$^{1}$}

\address{$^{1}$ ELI Beamlines Facility, The Extreme Light infrastructure ERIC, Za Radnici 835, 25241 Dolni Brezany,
	Czech Republic}
\address{$^{2}$ Department of Physics, University of Oxford, Clarendon Laboratory, Parks Road, Oxford OX1 3PU, United Kingdom}
\address{$^{3}$ York Plasma Institute, Department of Physics, University of York, Heslington, York, North Yorkshire YO10 5DD, UK}
\address{$^{4}$ Kansai Institute for Photon Science (KPSI), National
	Institutes for Quantum Science and Technology (QST), 8-1-7~Umemidai, Kizugawa-shi, Kyoto 619-0215, Japan}

\ead{Martin.Matys@eli-beams.eu}
\vspace{10pt}
\begin{indented}
\item[]\today
\end{indented}
\begin{abstract}
	The interaction of a high-power laser with a solid target provides ways to produce beams of $\gamma$-photons. For normal incidence of the laser on the target the beams usually appear in a form of two lobes, which are symmetric with respect to the laser propagation axis. In this work we demonstrate via three-dimensional particle-in-cell simulations a regime where for oblique incidence the emission of a collimated $\gamma$-photon beam is in the direction parallel to the target surface. The process is ascribed to the interference pattern in the electromagnetic field formed by the incident and reflected laser pulse. The electromagnetic field accelerates electrons to the GeV energy level, while temporarily directing their momentum along the target surface. Consequently, they emit a collimated $\gamma$-photon beam in the same direction. The dependencies of $\gamma$-photon emission on the incident angle, laser pulse polarization, power and duration and target thickness and preplasma are also addressed in the paper. The beam directionality is important for designing future experiments. In addition, this setup causes the generation of high-order harmonics propagating along the target surface.
	
\end{abstract}
\newpage
%
%
%
%
%

\section{Introduction}

An enormous progress in the development of extremely high-power laser technology resulting in the construction of many multi-PW laser facilities has been made all over the world \cite{Danson2019}. As a result, there is a fast growth of interest towards building compact laser-based accelerators of charged particles \cite{Tajima1979,Esarey2009,Bulanov2002,Daido2012,Macchi2013,Bulanov2014,Passoni2019} and radiation sources \cite{Teubner2009,Krausz2009}. Recent experiments on laser electron acceleration demonstrated electrons accelerated up to the energy of 10 gigaelectron volts (GeV) within a distance of tens of centimeters \cite{Gonsalves2019,Aniculaesei2023}.  In ion acceleration, proton beams with energy reaching 150 MeV have been demonstrated in the interaction of a laser with a thin foil target \cite{Ziegler2024}. High-power lasers are used for studying high-brightness high-energy photon sources \cite{Bulanov2013} and high-order harmonic generation (HHG) in the laser interaction with gas \cite{Pirozhkov2017_Biser} and solid density targets  \cite{Teubner2009, Lamaifmmodecheckcelsevcfi2023}. These lasers pave the way towards studying the previously unexplored domain of parameters corresponding to superstrong field for the quantum regimes of electromagnetic wave interaction with charged particles and the vacuum \cite{Mourou2006,Marklund_Shukla_2006, DiPiazza_Keitel2012,Gonoskov2022}.  These processes have been demonstrated through the  electron-positron ($ e^- $--$ e^+$) pair creation from vacuum in the seminal experiment \cite{Burke1997}. A further experiment has been recently proposed for proving the nonlinear property of vacuum \cite{Abramowicz2021}. It  was also predicted theoretically that through this interaction the $\gamma$-flash can be generated via multiphoton Compton scattering  with high conversion efficiency \cite{Ridgers2012,Nakamura2012,Lezhnin2018,Hadjisolomou2023_review} and the first experimental evidence for this has been recently presented \cite{Pirozhkov_2024_arxiv}. The multiphoton Compton scattering is characterized by the Lorentz invariant parameter $ \chi_e = \gamma_e\sqrt{\left(\textbf{E}+\textbf{v}\times \textbf{B}\right)^2-\left({\textbf{v}\cdot \textbf{E}/c}\right)^2}$, where $ \gamma_e $ is the Lorentz
factor of the electron with velocity $ \textbf{v} $, $ c $ is the speed of light in vacuum, and $ \textbf{E} $ and $ \textbf{B} $ are the laser electric and magnetic
fields normalized to the Schwinger field, $ E_S \approx 1.3\times10^{18}$ $ \mathrm{V/m}$. The $\gamma$-flash radiation has exciting applications in various fields, such as materials science at extreme energy densities \cite{Eliasson2013}, $\gamma$-ray inspection and imagining \cite{Albert_2016}, photonuclear
	reactions \cite{Ledingham2000_Photonuclear_Phys,Nedorezov2004,Kolenaty2022}, neutron sources \cite{Pomerantz2014_neutron}, photonuclear fission \cite{Cowan2000_fission,Schwoerer2003}, radiotherapy \cite{Weeks1997}, shock-wave studies \cite{Antonelli2017}, $ e^- $--$ e^+$ pair generation \cite{Ehlotzky2009} (which can be further guided using orthogonal  collision with a laser \cite{Maslarova2023, Martinez2023}) and in quantum technologies\cite{Ujeniuc_Suvaila_2024}, and it can be used to help understand the mechanisms of high-energy astrophysical processes \cite{Bulanov_Laboratory_Astrophysics_2015,Rees1992_fireball,Philippov2018_Pulsar,Aharonian_2021_Astrophysics}.

A typical approach towards realization of $\gamma$-photon radiation uses the interaction of a laser pulse with a solid density target \cite{Zhidkov2002,Koga2005,Gu_Klimo_Bulanov_Weber_2018,Hadjisolomou2023_review}. Here, the laser pulse accelerates electrons, which then interact with the same pulse.  Another setup to increase the $\gamma$-photon emission is to use two laser pulses \cite{Bell2008,Kirk2009,Luo2015,Grismayer2016,Hadjisolomou2025_wire} (or more of them \cite{Vranic2016,Gong2017}). The normal incidence of a linearly polarized laser on target results in a symmetric double-lobe structure of the emitted $\gamma$-photons, which is demonstrated in particle-in-cell (PIC) simulations for various cases \cite{Nakamura2012,Nerush2014,Stark2016,Wang2020_Double_Lobe,Wang2020,Vyskocil2020,Hadjisolomou2021,Hadjisolomou_2022_longer,Hadjisolomou2022,Galbiati2023,Formenti2024}.
However, in typical laser-solid interaction experiments, an oblique incidence angle is required to mitigate laser back-reflection which could damage the laser. In contrast to the double-lobe of the normal incidence, the structure and directionality of the emitted $\gamma$-photon beam(s) for oblique incidence varies strongly for different configurations. In addition to the incidence angle, the $\gamma$-photon directionality also depends on the target density, surface modulation \cite{Kleij2024} and target thickness, and various laser parameters (e.g., duration and focal spot size) \cite{Goodman_McKenna_2023}. Thus, the prediction of $\gamma$-photon beam direction becomes crucial for the reliable measurement of future experiments. 
	A possible solution is to use specific laser-target parameters to align the dominant direction of the emitted $\gamma$-photons to the target surface. This was previously reported using grazing laser incidence in combination with a relativistically near-critical target \cite{Serebryakov2016} (although limiting the $\gamma$-photon yield) and in studies on lower-energy radiations like extreme ultraviolet  (XUV) for oblique incidence (with a flat target \cite{Lamaifmmodecheckcelsevcfi2023} and a grating target \cite{Cantono_Macchi_2018}) and hard x-ray for parallel incidence \cite{Shen_Pukhov_Qiao_2024}. The use of a grating on the target surface or a very grazing laser incidence angle are connected with generation of surface plasma waves which can be used for electron acceleration \cite{Sarma_Macchi_2022} and HHG \cite{Cantono_Macchi_2018,Shen_Pukhov_Qiao_2024}. Moreover, for oblique incidence the interference between incident and reflected parts of the laser pulse generates a pattern in the electromagnetic field, accelerating electrons to high energies, as demonstrated via grazing incidence \cite{Chen2006,Serebryakov_2017_near_surface_el}. This approach can be extended to a non-grazing incidence and opaque target (where the generated $\gamma$-flash is separated from the reflected laser pulse) as demonstrated in this work. The electron momentum is temporarily directed along the target surface by the field pattern, which results in a collimated $\gamma$-photon beam being emitted in the same direction.
	
In this paper, we report the emission of a collimated $\gamma$-photon beam in the direction parallel to the target surface via fully three-dimensional (3D) PIC simulations for multiple laser and target parameters using non-grazing laser incidence angles and metallic (relativistically over-critical) target. 

The paper is structured as follows. The theory of electron dynamics in the interference pattern of the electromagnetic field is described in Section \ref{sec_theo}.  The simulation method and parameters are described in Section \ref{sec2}. The results are presented in Section \ref{sec_results}, which is divided into two subsections. In subsection \ref{sec31} we compare the normal incidence with 45\cc oblique incidence and describe the mechanism behind the collimated $ \gamma $-photon emission in the direction parallel to the target surface. This section also includes a Fourier transform analysis to observe the HHG. The dependence of the $\gamma$-photon emission on the laser incidence angle is shown in subsection \ref{sec_angles}. Section \ref{sec_conclusion} contains the discussion of the results and the conclusion. The Appendix describes the dependency of $\gamma$-photon emission on laser polarization, pulse duration, laser power, target thickness and preplasma and the use of Virtual Reality to visualize our data.

\section{Electron dynamics in the interference pattern of the electromagnetic field}\label{sec_theo}
When a p-polarized (the electric field oscillates in the plane of incidence) laser pulse interacts with an overdense target, an interference pattern is created in front of the target surface by the combination of the incident and the reflected parts of the laser pulse. The interference of the electromagnetic fields can be described by the superposition of an incident and a reflected planar wave \cite{Serebryakov_2017_near_surface_el}:

\begin{equation}\label{Ex}
	E_{\parallel}=2 E_0 \cos \theta_n \sin (k r_{\perp} \cos \theta_n) \sin \left(k r_{\parallel} \sin \theta_n -\omega t+\phi_0\right),
\end{equation}

\begin{equation}\label{Ey}
	E_{\perp}=2 E_0 \sin \theta_n \cos (k r_{\perp} \cos \theta_n) \cos \left(k r_{\parallel} \sin \theta_n-\omega t+\phi_0\right),
\end{equation}
\begin{equation}\label{Bz}
	B_{\odot}=2 E_0 \cos (k r_{\perp} \cos \theta_n) \cos \left(k r_{\parallel} \sin \theta_n-\omega t+\phi_0\right).
\end{equation}

Here, $ \parallel $ and $ \perp $ denote the directions parallel and perpendicular to the target surface inside the plane of incidence and $ \odot $ is the direction normal to this plane, $ r_\parallel $ and $r_\perp $ are the spatial coordinates in these directions respectively, $ k = 2\pi c/\lambda $ is the laser wavenumber, $\lambda $ is the laser wavelength, $ E_0 $ is the field amplitude, $ \phi_0 $ is the initial phase and $ \theta_n $ is the incidence angle (the angle between the target normal and the laser axis).

The interference pattern corresponds to a wave with superluminal phase velocity, $ v_{ph} = c/\sin(\theta_n)>c $,  propagating along the target surface  and with wavelength $\lambda_{\perp}=\lambda/\sin(\theta_n)$. In the perpendicular direction the field pattern repeats with a spatial period $ \Lambda_{\perp}=\lambda/\cos(\theta_n)$. Thus, a position $ r_{\perp}=\Lambda_{\perp}/4$ exists in front of the target where the terms $\sin (k r_{\perp} \cos \theta_n) = 1 $, $\cos (k r_{\perp} \cos \theta_n) = 0$ and equations \ref{Ex}-\ref{Bz} simplify to:

\begin{equation} \label{eq:e_long_simpl}
	E_{\parallel}=2 E_0 \cos \theta_n \sin \left(k r_{\parallel} \sin \theta_n-\omega t+\phi_0\right),\qquad E_{\perp}=0,\qquad B_\odot=0.
\end{equation}
Therefore, around this position a bunch of electrons is accelerated along the target surface by the electric field $ E_{\parallel} $. As the wave in the interference pattern is superluminal, an electron is accelerated only during the time that it is in phase with the wave, i.e., when it is moving in the negative half of its sinusoidal profile. Thus, the maximal phase difference is $ \pi $. By using the time averaged field (from 0 to $ \pi $) from eq. \ref{eq:e_long_simpl}: $ \overline{E_{\parallel}} = 4\cdot E_0\cos\theta_n/\pi $, Ref. \cite{Serebryakov_2017_near_surface_el} gives the maximal electron Lorentz factor as 

\begin{equation}\label{eq:gamma:max}
	\gamma_{max} = 1 + \frac{4 a_0\cos \theta_n}{1 - \sin \theta_n},
\end{equation}
where $ a_0 $ is the normalized dimensionless electromagnetic wave amplitude. Note that in the theory used in Ref. \cite{Serebryakov_2017_near_surface_el}  the incidence angle is defined as the angle between the laser propagation axis and the target surface. Thus, $\cos \theta_n$ and $\sin \theta_n$ are swapped in Ref. \cite{Serebryakov_2017_near_surface_el} compared to this work.  

Our laser field has a Gaussian spatiotemporal profile (in contrary to the plane wave used above) and thus the locations where $ E_\perp =0 $ and $ B_\odot =0 $ do not form straight lines. Therefore, the acceleration time is shorter \cite{Serebryakov_2017_near_surface_el}. Also in PIC simulations the laser reflection on the target is imperfect and the target bends under the radiation pressure. Nevertheless, $ \gamma_{max} $ in our simulations is of the same order of magnitude as in Eq. (\ref{eq:gamma:max}) and its dependency on the incidence angle is the same as in the theoretical model.

The electrons obtain higher maximum energy for larger incidence angles \cite{Serebryakov_2017_near_surface_el},  as $ \gamma_{max} $ in Eq. (\ref{eq:gamma:max}) tends to infinity  with $ \theta_n  $ approaching $ \pi/2 $. On the other hand, $ E_\parallel $ becomes stronger with decreasing incidence angle and is proportional to $2\cos \theta_n $. Therefore, the use of smaller angles is beneficial for $\gamma$-photon directionality, as shown in this paper. The strong localized field provides rapid electron acceleration over a short distance (for grazing angles the transverse drift is significant over larger distances, $ L >> \lambda $ \cite{Serebryakov_2017_near_surface_el}).  This causes $ \gamma $-flash emission predominantly in the direction parallel to the target surface, since the direction of the emitted photons coincides with that of the emitting electrons (as $\gamma_{e}>>1$). 

\section{Simulation setup and parameters} \label{sec2}
\begin{figure}[ht]
	\begin{center}
		\flushleft
		\includegraphics[width=1\linewidth]{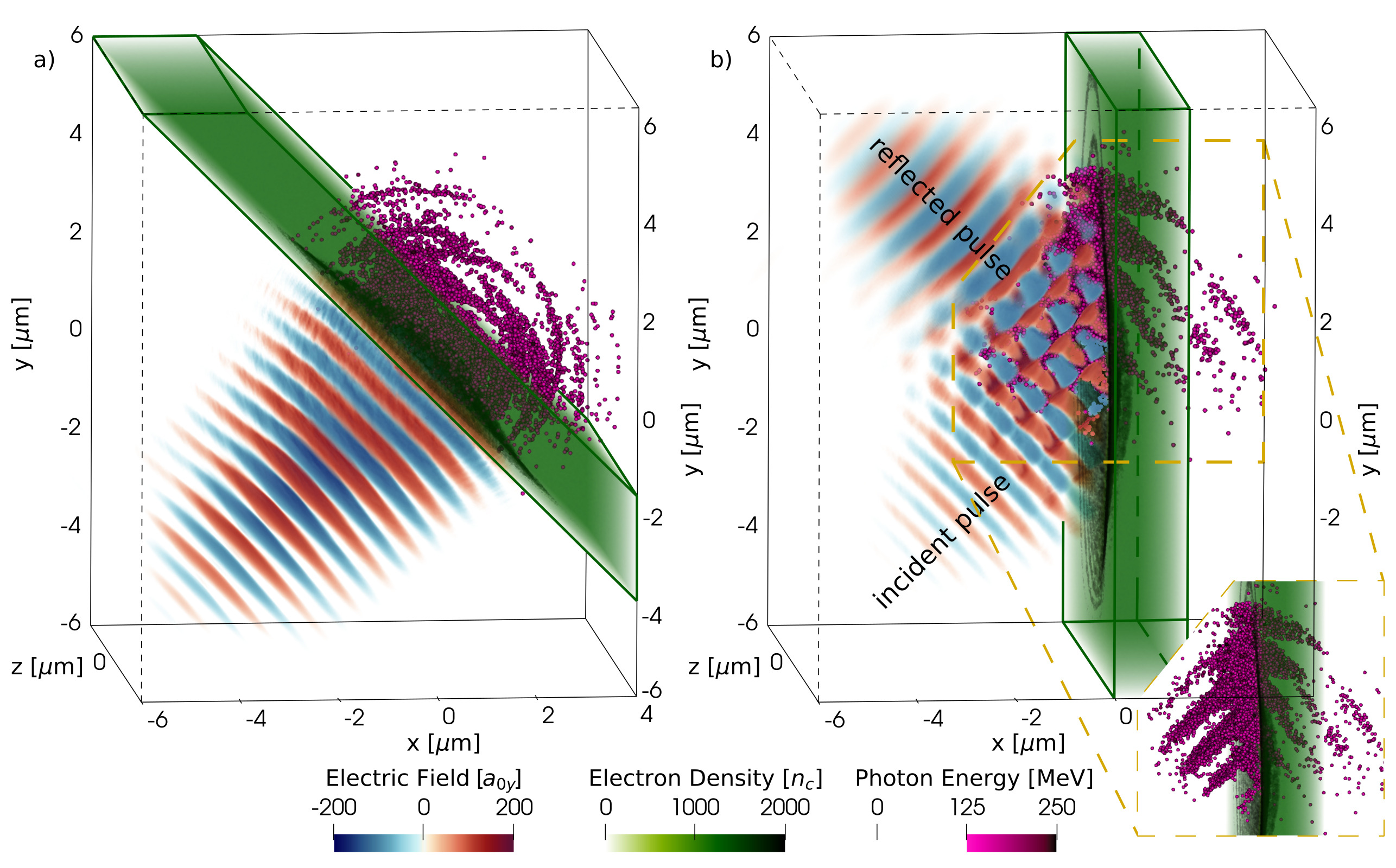}	
		\caption{\label{fig:sim_setup} Simulation setup at 67.5 fs. a) normal incidence, b) 45$^\circ$ incidence. The inset shows the highlighted area without the laser.
			The target is visualized as transparent to show the $\gamma$-photons, while the edges of the target are highlighted. The color scales are saturated at the displayed values. The electric field in the $ y $-direction is presented in the units $ a_{0y}= eE_y/m_e\omega c $.}
	\end{center}
\end{figure}

The 3D simulations were done with the PIC code EPOCH \cite{Arber2015}, including the Higuera-Cary pusher and QED module for non-linear Compton scattering. The Bremsstrahlung process \cite{KOCH1959,Vyskocil2018_Brems} is omitted in this work, as Compton scattering dominates over it at ultra-high laser intensities for times comparable to the laser pulse duration. The simulation box size is 14.336 $ \mu $m $\times$ 20.48 $ \mu $m $\times$ 10.24 $ \mu $m, the cell size is 4 nm $\times$ 4 nm $\times$ 32 nm and 4 macro-electrons and 4 macro-ions are
assigned to each cell containing plasma. A convergence test using double number of particles has been made with minimal changes observed.    
 In all simulations the laser propagates from the boundaries at positions $ x = y = -10.24 $ $ \mu $m at an angle of 45$^\circ $ (the plane of incidence is the $ x $-$ y $ plane), as shown in figure \ref{fig:sim_setup}. The incidence angle is determined by the target orientation, as shown in Fig. \ref{fig:sim_setup}(a) for the  normal incidence and \ref{fig:sim_setup}(b) for the 45$^\circ $ incidence.   
 
The reference laser is linearly p-polarized. The laser wavelength is $ 0.8\ \mathrm{\mu m}$ and the peak intensity is $ I_{\mathrm{max}} = 2\times10^{23}\  \mathrm{W/cm^2}$, thus $ a_0 \approx 0.85\sqrt{I\left[10^{18}\mathrm{W\ cm^2}\right]\lambda^2\left[\mathrm{\mu m}\right]}  \approx 304 $. The critical plasma density is $n_c =~m_e \omega^2/4\pi e^2 \approx 1.742\times10^{21}\  \mathrm{cm^{-3}}$, where $ \omega $ is the laser angular frequency, $ m_e $ is the electron mass and $ e $ is the electron charge. The laser pulse has Gaussian spatial and temporal profiles with beam width equal to 3 $\lambda $ (2.4 $ \mu $m) at the full width at half maximum (FWHM) and duration of 20 fs at FWHM (in intensity). The total laser energy for the reference case is 283 J in the 3D simulation, thus the laser power is approximately 14 PW. The target is made of fully ionized iron with an electron number density of $ 2.22\times10^{24}\ \mathrm{cm}^{-3}$ (about 1274 $ n_c $).  The target thickness is 1.5 $ \mu $m.

 The QED module in EPOCH code includes the recoil effect on electrons where they lose momentum equal to the momentum of the emitted photons. Moreover, it  provides options to include photons dynamics after they are generated and the production of $ e^- $--$ e^+$ pairs. Aside of the possible pair production, the photons currently do not further interact with the plasma or electromagnetic fields and propagate ballistically after being generated. However, the effect of pair production is not significant for our simulation parameters as only one positron macroparticle (corresponding to 284160 positrons) is generated for the 45\cc case (in comparison $ 2.5\times10^9 $ electron macroparticles of the same numerical weight is initialized in the simulation) and 15 positron macroparticles (corresponding to $ 4\times 10^6 $ positrons) for the 0\cc case of simulations shown in Fig. \ref{fig:sim_setup}. Therefore, we disable photons dynamics and the pair production afterward to reduce the computational time.  Furthermore, only $ \gamma $-photons and electrons with energy above 1 MeV are stored for evaluation (photons with lower energy are still generated  and contribute to the electron recoil). The visualization of the $\gamma$-photon emission temporal evolution of the 45\cc case  can be accessed online \cite{VBL_mm_gamma}  via our Virtual Beamline -- VBL application \cite{VBL_home_page,Danielova2019,Matys2023}, described at the end of the Appendix.  
\section{Results}\label{sec_results}

\subsection{Generation of collimated $ \gamma $-photon beam along the target surface} \label{sec31}

When the laser polarization lies in the laser incidence plane then the angle between the target and the incident laser influences the directionality of the emitted $\gamma$-photons, as already shown in Fig. \ref{fig:sim_setup}. For the normal incidence case (fig. \ref{fig:sim_setup}-a) the $\gamma$-photons propagate mainly in the approximate forward direction, forming a double-lobe structure. Moreover, they are emitted periodically, with the period matching that of the laser. The situation changes for the oblique incidence (45$^\circ$ in fig. \ref{fig:sim_setup}-b) where the $\gamma$-photons move predominantly in the direction parallel to the target surface and partially in the incident and reflected laser directions, as seen in the inset of Fig. \ref{fig:sim_setup}-b. In the rest of the paper, only simulations with immobile photons are presented. 

\begin{figure}[ht]
	\begin{center}
		\flushleft
		\includegraphics[width=1\linewidth]{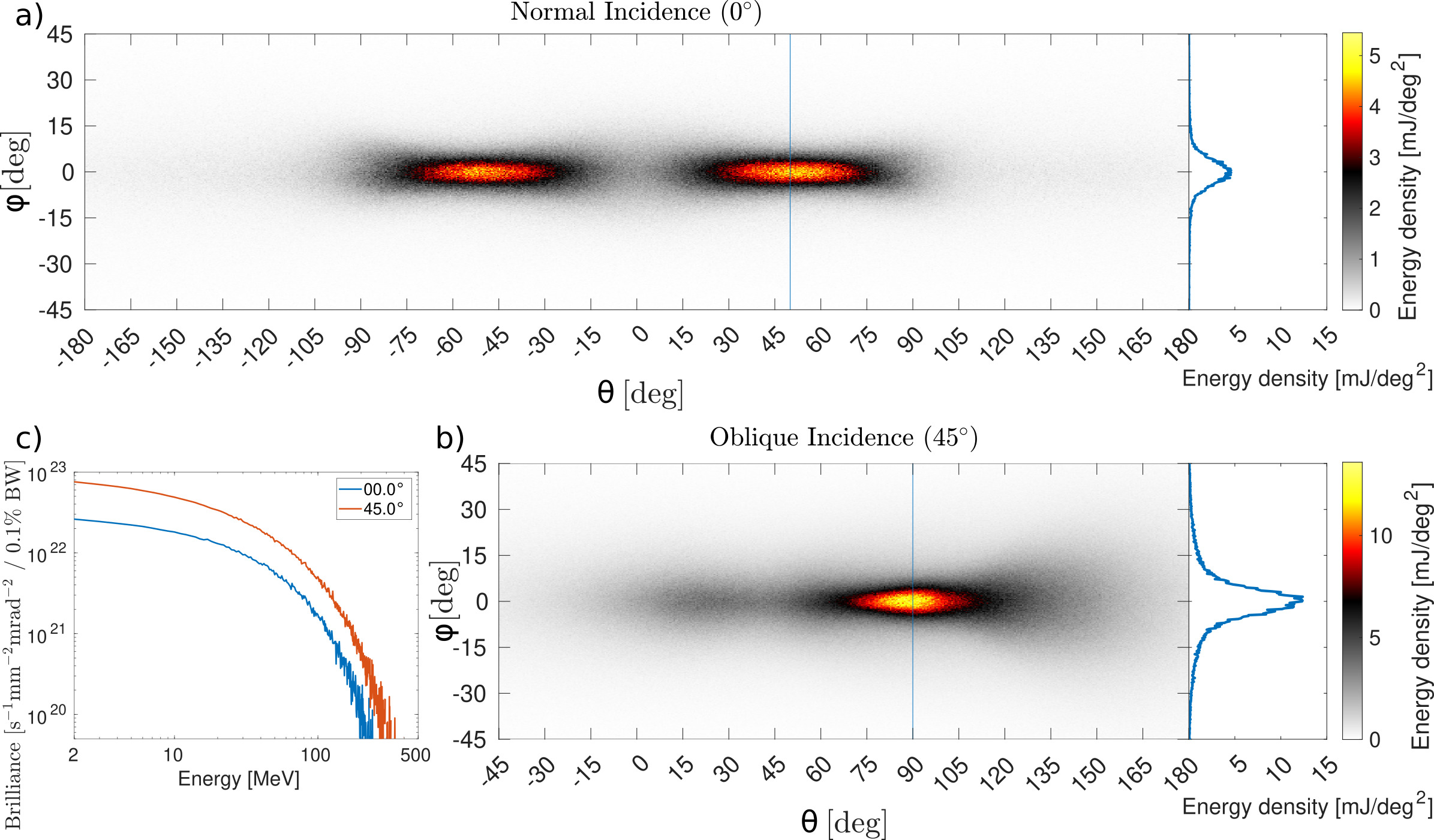}	
		\caption{\label{fig:spherical_distribution_00x45} 3D angular energy distribution of the emitted $ \gamma $-photons  at the end of the simulation for a) 0$ ^{\circ}$ and b) 45\cc incidence. The lineouts show the energy distribution in the transverse direction taken at angles $ 90^\circ$ (45\cc case) and $ 50^\circ$ (0\cc case). c) Brilliance for these cases at the same angles. }
	\end{center}
\end{figure}

The $\gamma$-photon directionality is shown in Fig. \ref{fig:spherical_distribution_00x45} via the angular distribution of the emitted $\gamma$-photon energy in the 3D space. Due to the setup geometry with the $ x $-$ y $ polarization plane (whose 2D slices at $ z=0 $ are further examined later), it is convenient to represent the angular distribution using the angles $ \theta = \mathrm{arctan}(p_y/p_x) $ and $ \phi = \mathrm{arctan}(p_z/p_r) $  in Fig. \ref{fig:spherical_distribution_00x45}, where $ p_{x,y,z} $ are the momenta in the respective directions of the Cartesian system of the simulation and $ p_r = \sqrt{p_x^2+p_y^2} $. Note that as targets are tilted by a different angle for different laser incidence angles, the angular data are shifted in the $ \theta$-direction by an angle equal to the angle between the target and the $ y $-axis of the simulation. The normal incidence case (Fig. \ref{fig:spherical_distribution_00x45}-a) creates a typical double-lobe structure. For our simulation parameters the maximum energy density is observed around $ \pm 50^\circ$. The situation differs for the 45$^\circ$ case (figure \ref{fig:spherical_distribution_00x45}-b) where a single spot around 90$^\circ$ (direction parallel to the target surface) is observed.  The lineouts of the blue lines (right part of Fig. \ref{fig:spherical_distribution_00x45}) show the energy distribution in the transverse direction at angles $ 50^\circ$ and $ 90^\circ$ for the normal and 45\cc incidence cases respectively. In both cases the $\gamma$-photon beam is well collimated in the transverse direction (angular spread at FWHM is around 10\cc for both cases) but the maximal energy density is more than double in the case of oblique incidence. The brilliance in Fig. \ref{fig:spherical_distribution_00x45}-c is calculated at the direction of the peak $\gamma$-photon yield (at the same angles as the lineouts) using an opening full-angle of 10\ccnd. Our simulations reveal a high brilliance for the 45\cc case, similar to that of Ref. \cite{Hadjisolomou2022}, approaching $ 10^{23} $ s$^{-1}$mm$^{-2}$mrad$^{-2}$ per 0.1\% bandwidth in the tens of MeV $\gamma$-photons energy range. In comparison, the brilliance for the normal incidence case is approximately three times lower, which is reflected to the ratio of the conversion efficiency, as discussed in Section \ref{sec_angles}.  

The $\gamma$-photon directionality depends on the structure of the electromagnetic field where the electrons are accelerated. In the case of oblique incidence, the incident and reflected laser pulses generate an interference pattern as seen in Fig. \ref{fig:sim_setup}-b, drastically different than the field in the normal incident case.

\begin{figure}[ht]
	\begin{center}
		\flushleft
		\includegraphics[width=1\linewidth]{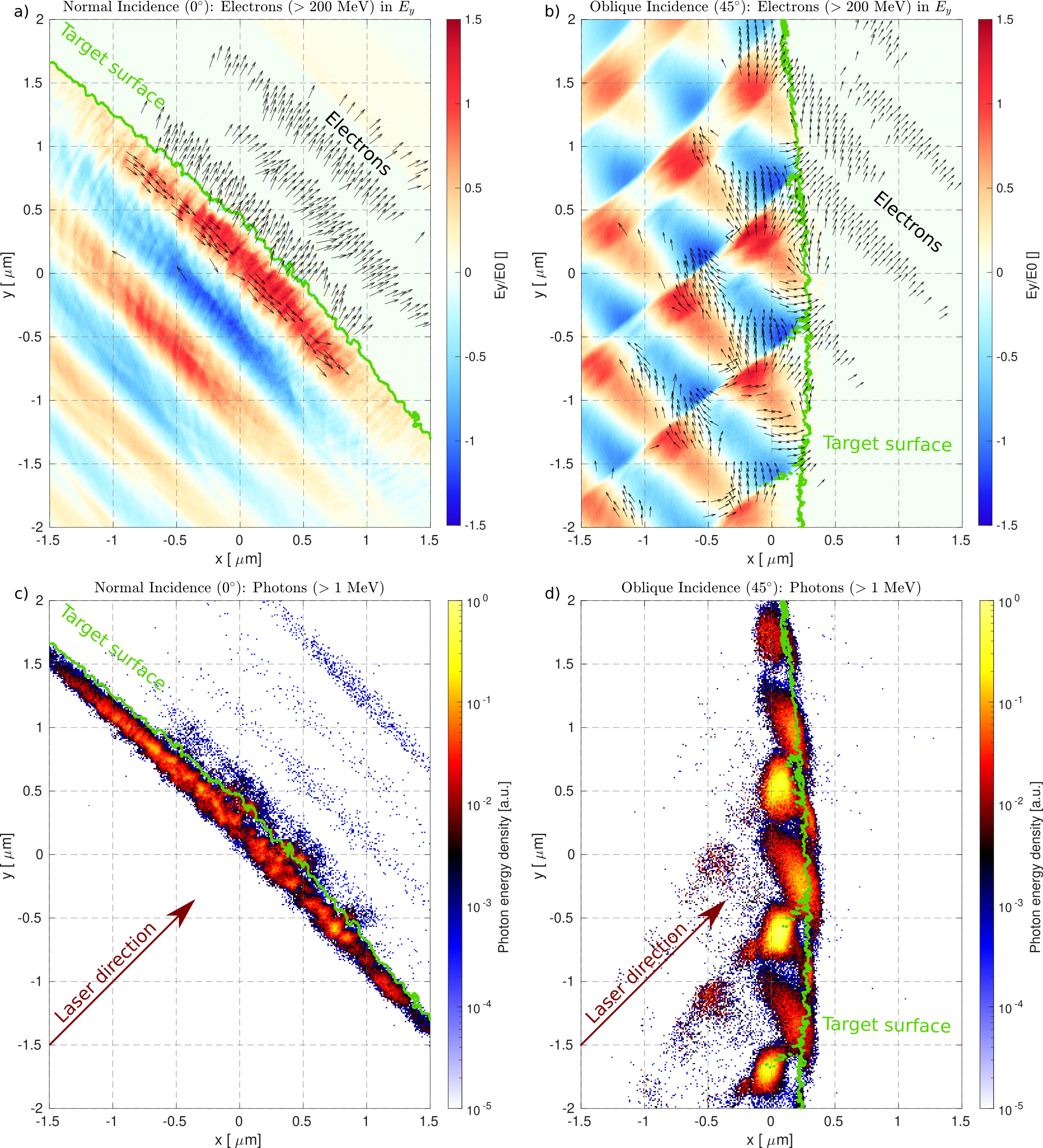}	
		\caption{\label{fig:field_electron_gamma} a-b) Structures of $ E_y$ at 60 fs. $ E_0 $ is the initial laser peak electric field. The black arrows show the average electron momentum vectors. The green contours depict the instantaneous target surface. c-d) Energy density of $\gamma$-photons emitted during the time interval of 59.66 -- 60 fs. a,c) normal incidence, b,d) 45$ ^{\circ} $ incidence.}
	\end{center}
\end{figure}

For further analysis 2D slices of the 3D fields and 2D angular distributions of photon energy around the polarization plane are used. Only particles with
\begin{equation}\label{eq:conditionz}
	\frac{\left|p_z\right|}{\sqrt{p_x^2+p_y^2}} =\frac{\left|p_z\right|}{{\left|p_r\right|}} < \tan\left(5^\circ\right)
\end{equation}
are considered in the 2D ($ x $-$ y $ plane) analysis of the simulation data, which corresponds to the $\gamma$-photon beam width at the FWHM in the transverse direction (see Fig. \ref{fig:spherical_distribution_00x45}).
 The $ E_y $ slices on the $ x $-$ y $ plane ($ z=0 $) are shown in Fig. \ref{fig:field_electron_gamma}-a,b, where the arrows depict the direction of the average electron momentum (for electrons with energy above 200 MeV, inside squares of 60 nm $ \times $ 60 nm) and the green lines depict the instantaneous target front surface (where the density equals the initial electron density).

 For normal incidence (fig. \ref{fig:field_electron_gamma}-a,c), electrons are accelerated towards one side of the target during the first half laser period and towards the other in the second one (fig. \ref{fig:field_electron_gamma}-a). Therefore, $\gamma$-photon emission (shown in fig. \ref{fig:field_electron_gamma}-c) occurs mostly on the target surface, as the electrons rotate to the opposite direction (when the field switches sign). 

  For the $ 45^\circ $ laser incidence case (shown in Fig. \ref{fig:field_electron_gamma}-b,d) $\gamma$-photons are emitted by three distinct mechanisms, occurring at different locations and with different electron trajectory patterns. 
  
  Firstly, as described in Section \ref{sec_theo}, the reflected laser pulse interferes with the incident one and creates an interference pattern, which propagates with superluminal phase velocity along the target surface (as seen in Fig. \ref{fig:field_electron_gamma}-b). Inside this pattern, electrons are accelerated to energies several times higher than that given by the quiver velocity \cite{Serebryakov_2017_near_surface_el}. As the field pattern moves faster than electrons, they experience a strong acceleration (and subsequent deceleration when the pattern changes sign) while temporarily aligning their momentum with the target surface. The direction of motion of the emitted $\gamma$-photons is approximately the same as that of the emitting electrons. Therefore, directional $ \gamma $-photon emission occurs at the field extremes, as in Figs. \ref{fig:field_electron_gamma}-b,d. The field extremes and the regions of high $\gamma$-photon energy density appear at $ y = [-1.7, -0.6, 0.55, 1.7] $ $ \mu\mathrm{m} $, with their spacing as predicted by the equation  $\lambda_{\perp} = \lambda/\sin \theta_n = 0.8\ \mu\mathrm{m}/\sin(45^\circ)\approx 1.13\  \mu\mathrm{m}$. These spots appears around $ x=0 $, which at time 60 fs
  is a low-density plasma region in front of the target (as the target surface shifts inwards under the radiation pressure). The instantaneous target surface is shown by the green contour in Fig. \ref{fig:field_electron_gamma}. 
  
  Secondly, a weaker $\gamma$-photon emission occurs at the locations between the highest $\gamma$-photon energy density spots, appearing  as prolonged structures centered at $ y=[-1.25,-0.25,0.75] $ $ \mu\mathrm{m} $ and close to the instantaneous target surface, as seen in Fig. \ref{fig:field_electron_gamma}-d. In these areas, a fraction of electrons (see Fig. \ref{fig:field_electron_gamma}-b) propagate towards the target, in a similar way as in the normal laser incidence case (shown in Fig. \ref{fig:field_electron_gamma}-a). Thus fewer $\gamma$-photons are emitted between approximately 0\cc and 45\ccnd, as shown in Fig. \ref{fig:spherical_distribution_00x45}-b.
  
  Thirdly, with an even smaller contribution, a $\gamma$-photon population is emitted from the instantaneous target surface ($ y=[-1, 0.25] $ $ \mu\mathrm{m} $ at 60 fs). There, electrons propagate along the target surface with wiggling movement, as seen in Fig.  \ref{fig:field_electron_gamma}-b.  Note that in our case the electron population propagating along the target surface (as in the surface plasma wave \cite{Cantono_Macchi_2018,Sarma_Macchi_2022,Shen_Pukhov_Qiao_2024}), moving outside of the interference pattern region ($ y\approx\pm 3 \mu$m)  does not significantly contribute to the $\gamma$-photon emission  as the $\chi_e$ factor for those electrons is small.
  
\begin{figure}[ht]
	\begin{center}
		\flushleft
		\includegraphics[width=1\linewidth]{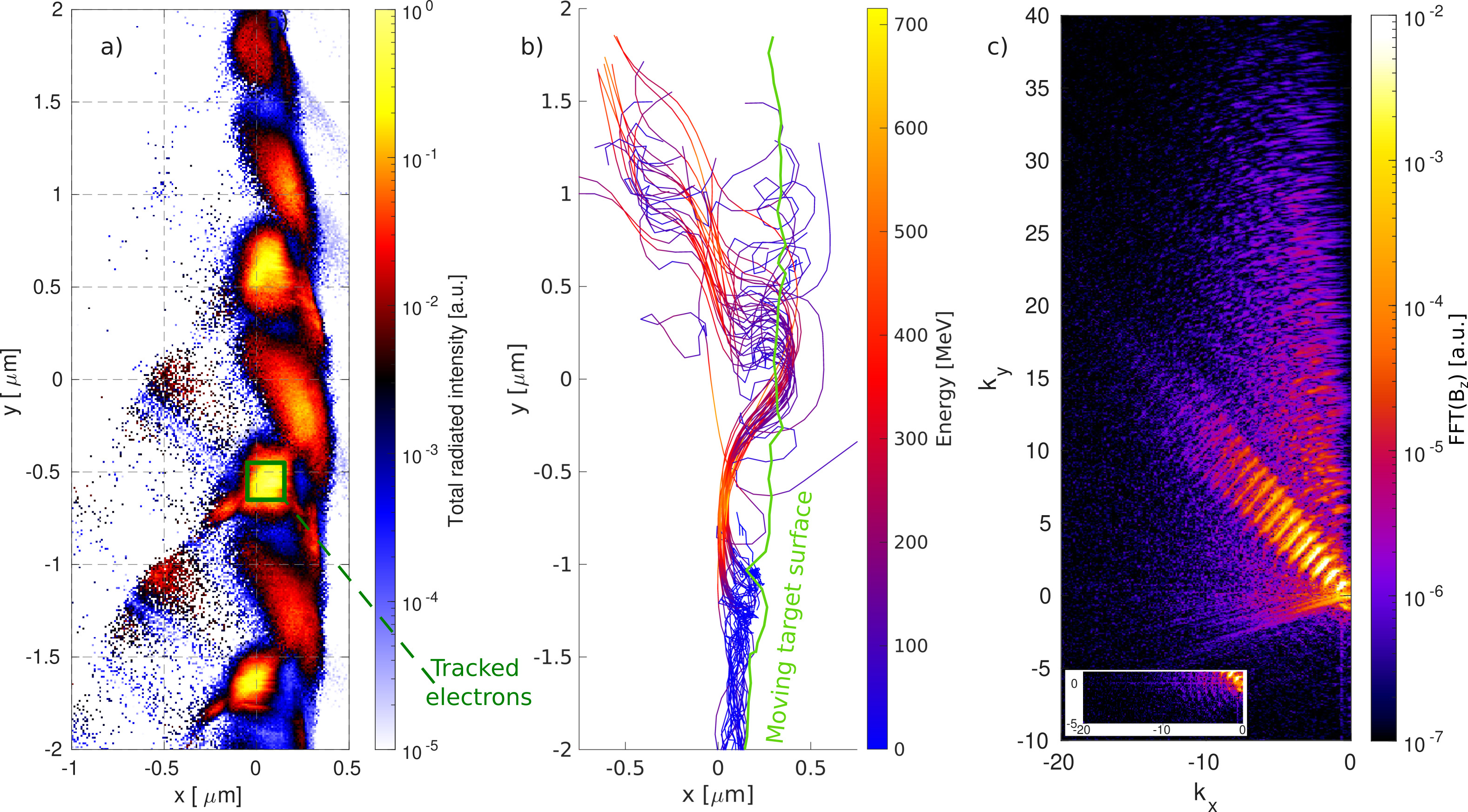}	
		\caption{\label{fig:trajectory} a) Total radiated intensity for the 45$^\circ$ at 60 fs. b) Trajectories of the 34 macro-electrons with the highest $ \chi_e $ in the spot highlighted by the green rectangle in fig. a). The green line in b) depicts the evolving (with respect to the $ y $-position) front target surface. c) 2D fast Fourier transform (FFT) of $ B_z $ at $ t= 90 $ fs. The inset shows the disappearance of the straight lines when  $ B_z $ at the target position and beyond is set to zero.} 
	\end{center}
\end{figure}

  Since the $\gamma$-photon emission from the first mechanism dominates, the electron trajectories from one of the bright spots is further investigated. Fig. \ref{fig:trajectory}-a is produced from the electron data by calculating the total radiated intensity ($ I_R $) of  nonlinear Compton scattering for every electron (with energy above 1 MeV and fulfilling the momenta condition in eq. \ref{eq:conditionz}) using the integral (equation 4.50 in Ref. \cite{Baier1998}):  
  \begin{equation}\label{eq:integral_RI}
  	I_{R} = C\int_{0}^{\infty}\frac{u\left(4u^2 + 5u + 4\right)}{\left(1+u\right)^4} K_{2/3}\left(\frac{2u}{3\chi_e}\right)\mathrm{d}u
  \end{equation} 
Here, $ C = e^2m_e^2/3\sqrt{3}\pi\hbar^2$, where $ \hbar $ is normalized Planck constant, $ K_{2/3} $ is the Modified Bessel function of the second kind and $ u $ is the integration variable. Note that Fig. \ref{fig:trajectory}-a is similar to Fig. \ref{fig:field_electron_gamma}-d (although Fig. \ref{fig:trajectory}-a is more clear since the electron data are instantaneous).

 Next, we tracked electrons with $ \chi_e \ge 0.3$ (34 macroparticles), located in the bright spot area shown in Fig. \ref{fig:trajectory}-a. The electron trajectories are shown in Fig. \ref{fig:trajectory}-b, where their energy is represented by the color scale (time evolution of the location of selected particles in $ B_z $, $ E_x $, $ E_y $ is shown in the supplementary video).
 The evolving (with respect to the $ y $-position) front target surface is depicted by the green line, the $ x $-position represents the first position with electron number density equal or higher than the initial target density at the corresponding $y$-position and  $ z=0 $, where the $ y $-positions are linked to one of the tracked electrons. The tracked electrons begin with a wiggling movement close to the front target surface as they reflect from the strong quasi-static magnetic field (generated by the space charge inhomogeneity). Afterwards, they are injected into the field interference pattern, align parallel to the initial target surface, and are accelerated to energies up to 700 MeV. Consequently, they radiate during the acceleration and deceleration processes around the spot position (approximately $ y=-0.5 $ $ \mu\mathrm{m} $) with $\gamma$-photons emitted in the direction parallel to the target surface.  Afterwards, the electrons mostly return towards the target surface. Periodically, another group of electrons are injected into the superluminal interference pattern and the process is repeated.
 
Fig. \ref{fig:trajectory}-c shows the 2D Fourier transform of $B_{z}$ slice (at $ z=0 $) at a time of $ 90 $ fs. The figure shows the HHG along the specular direction due to the mechanism of relativistic oscillating mirror (ROM) \cite{Bulanov1994_ROM} (see also review article \cite{Teubner2009}
and references cited therein). Furthermore, the figure shows the recently discovered XUV relativistic instability-modulated emission (RIME) propagating along the surface-parallel direction \cite{Lamaifmmodecheckcelsevcfi2023}. The RIME process is driven by a strong quasi-static magnetic field at the target surface which is generated by the oblique incidence \cite{Grassi2017}. This field confines part of the electrons near the plasma surface and prevents them from passing through it. In turn, oscillating relativistic nanobunches are generated via the two-stream instability, which are responsible for the surface-parallel XUV emission seen in Fig. \ref{fig:trajectory}-c, and which also help with injection of electrons from the target surface into the interference pattern in front of it. The presence of the quasi-static field is shown by the straight lines in Fig. \ref{fig:trajectory}-c in the area $ k_y < 0 $. The straight lines disappear from the Fourier transform result when $ B_z $ at the target position and beyond is artificially removed from the data, as is shown in the inset of Fig. \ref{fig:trajectory}-c.

\begin{figure}[ht]
	\begin{center}
		\flushleft	
		\includegraphics[width=1\linewidth]{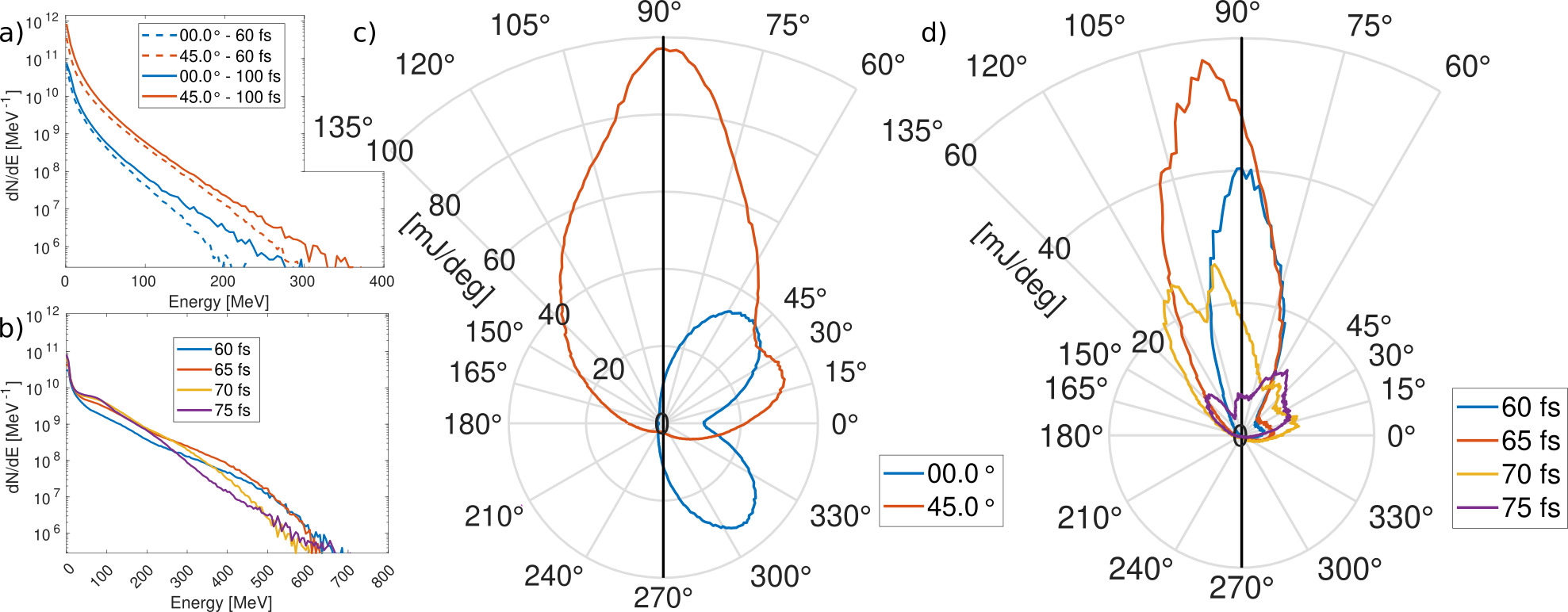}
		\caption{\label{fig:spectra_rate} a) $\gamma$-photon energy spectra comparing normal (blue) and 45$^\circ$ (red) incident angles at 60 fs (dashed lines) and at 100 fs (solid lines). b) Electron energy spectra of the 45\cc case at different times, shown in the figure legend. c) 2D angular distribution of the $\gamma$-photon energy at 100 fs. d) 2D angular distribution of electron energy using electrons with $ \chi_e>0.1 $. }
	\end{center}
\end{figure}

The $\gamma$-photon energy spectra at time of 60 fs (the time used in Figs. \ref{fig:field_electron_gamma} and \ref{fig:trajectory}(a)) and at 100 fs (end of the simulation) for normal and 45\cc incidence are compared in Fig. \ref{fig:spectra_rate}-a.
The 45\cc incidence case produces approximately one order of magnitude more $\gamma$-photons compared to the normal incidence case. The $\gamma$-photon cut-off energy for the 45\cc incidence case is above 350 MeV, compared to less than 300 MeV for the normal incident case (the bin size is 5 MeV and the minimal value of the dN/dE = 284160 MeV$ ^{-1} $ corresponds to five macroparticles in the 45\cc case). Fig. \ref{fig:spectra_rate}-c  shows the angular distribution of $\gamma$-photon energy in the $ p_y $/$ p_x $ plane, using $\gamma$-photons  fulfilling eq. \ref{eq:conditionz} as discussed above. The 45\cc incidence simulation predominantly emits $\gamma$-photons in the  direction parallel to the target surface  (at $ 90^\circ $, depicted by the black line), while the normal incidence case is characterized by the typical double-lobe pattern. Moreover, the 45\cc incidence case emits approximately 3 times more $\gamma$-photons at the 90\cc emission angle compared to the maximum of the normal incidence case. The time evolution of electron energy spectra is shown in fig. \ref{fig:spectra_rate}-b. The cut-off electron energy is almost double compared to the $\gamma$-photon energy, approaching 700 MeV. The number of high-energy electrons with energy above 300 MeV drops significantly between 65 fs and 70 fs when a high number of $\gamma$-photons is emitted. 
 The angular distribution of electron energy is shown in Fig. \ref{fig:spectra_rate}-d. Here, only electrons with $ \chi_e>0.1 $ are considered in the distribution. At the time of 60 fs the high $ \chi_e $ electrons have their momentum oriented at 90\cc (target surface). At later times, the momentum orientation of these electrons gradually drift away as the target is bended by the radiation pressure. Therefore, a broadening of the $\gamma$-photon angular energy distribution gradually occurs, as seen in Fig. \ref{fig:spectra_rate}-d.  At a time of 70 fs, the electron angular distribution separates into several directions. However, by this time their effect on the $\gamma$-photon distribution is becoming less important, as the total number of electrons with high $ \chi_e $ is decreasing. 

\subsection{Dependence on the laser incidence angle} \label{sec_angles}
\begin{figure}[ht]
	\begin{center}
		\flushleft
		\includegraphics[width=1\linewidth]{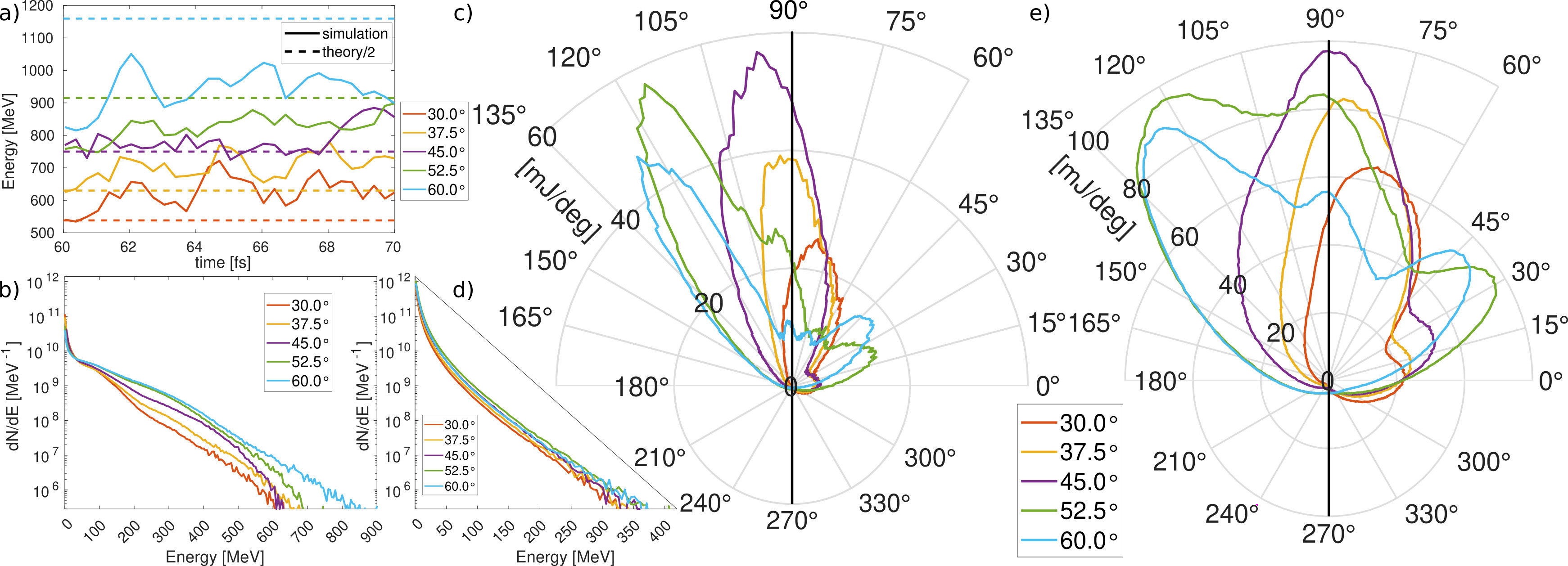}	
		\caption{\label{fig:angular_distribution_angles} All graphs share the same coloring based on the incident angle: a) Time evolution of the maximum electron energy. b) Electron energy spectra at 65 fs. c) 2D angular distribution of electron energy in the $ x $-$ y $ plane for electrons with $ \chi_e>0.1 $. d) $\gamma$-photon energy spectra at 100 fs. e) 2D angular distribution of $\gamma$-photon energy.} 
	\end{center}
\end{figure}

In this Section we compare simulations  with incidence angles ranging from 30$^\circ$ to 60$^\circ$. As predicted in Ref. \cite{Serebryakov_2017_near_surface_el}, the maximum energy of electrons accelerated in the interference pattern increases with the angle, as can be seen in Figs \ref{fig:angular_distribution_angles}-a,b.  
	The solid lines in Fig. \ref{fig:angular_distribution_angles}-a show the maximum electron energy at different simulation times. The dashed lines show the electron energy predicted by Eq. \ref{eq:gamma:max} (divided by a factor of two), which is in good agreement with our simulations. The model correctly predicts the increase of the maximum electron energy with incident angle. Fig. \ref{fig:angular_distribution_angles}-b shows the energy spectra at time of 65 fs, i.e., when the electron acceleration in the interference pattern is still ongoing. At a later time, the electrons are directly accelerated by the reflected laser pulse in the specular direction (see Ref. \cite{Thevenet_2015} and the references cited therein). However, the directionality of the electron beams gradually shifts away from the target surface when larger angles are used, as shown in Fig. \ref{fig:angular_distribution_angles}-c, where we consider electrons of $ \chi_e > 0.1 $.

The $\gamma$-photon energy spectra at 100 fs are shown in Fig. \ref{fig:angular_distribution_angles}-d. Both the number of emitted $\gamma$-photons and their cut-off energy increase with the angle, although this increase is less apparent than that of the electrons. Exceptionally, for the 52.5\cc incidence angle case more $\gamma$-photons are emitted than in the 60\cc case. As seen in Fig. \ref{fig:angular_distribution_angles}-c the 52.5\cc case develops an additional central lobe at 90\cc and the spectrum contains higher energy density of electrons with $ \chi_e > 0.1 $ than the 60\cc case. Consequently, the $\gamma$-photon energy angular distribution (Fig. \ref{fig:angular_distribution_angles}-e) of the 52.5\cc case contains more energy than the 60\cc case. Therefore, the interference pattern effect, described in Section \ref{sec31}, is utilized more efficiently in the 52.5\cc incident angle case than in the 60\cc case. 

Overall, the angular dependency of the $\gamma$-photon density distribution (Fig. \ref{fig:angular_distribution_angles}-e) shows a gradual shifting from the double-lobe structure, for the normal incidence case, to a single lobe along the target surface for the 45\cc case. The 30\cc and 37.5\cc cases reveal a smooth transition between the 0\cc and 45\cc cases. When increasing the incident angle above 45\ccnd, a third lobe is emitted near the laser specular direction. In addition, all our simulations exhibit a local minimum approximately in the direction of the incident laser pulse.   

\begin{figure}[ht]
	\begin{center}
		\flushleft
		\includegraphics[width=1\linewidth]{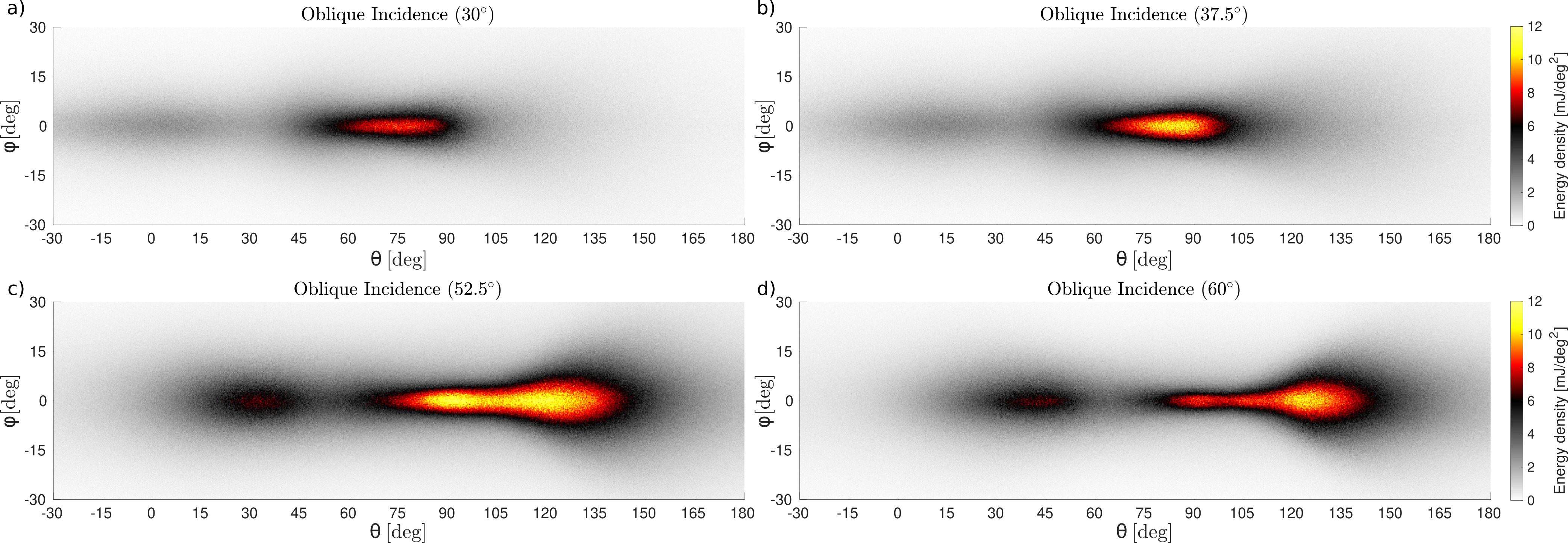}	
		\caption{\label{fig:spherical_distribution_other_angles} 3D angular energy distribution of the emitted $ \gamma $-photons  at the end of the simulation for different incidence angles: a) 30\ccnd, b) 37.5\ccnd, c) 52.5\cc and d) 60\ccnd.}
	\end{center}
\end{figure}

The 3D angular energy distribution of the emitted $ \gamma $-photons for these cases is shown in Fig. \ref{fig:spherical_distribution_other_angles} (the 0\cc and 45\cc cases are shown in Fig. \ref{fig:spherical_distribution_00x45}). 
 A spot at 90\cc occurs for the cases of 37.5\ccnd, 45\ccnd, 52.5\cc and 60\ccnd.

 The laser energy (283 J) to $\gamma$-photon energy conversion efficiency for various incidence angles is shown in Table \ref{table_en_conv}. The simulations with oblique incidence converted up to 3.5 times more energy into $\gamma$-photons than the simulation with normal incidence. The conversion efficiency rises with the angle till 52.5\ccnd, where it reaches 10.7 \%, and decreases afterwards. The rest of Table \ref{table_en_conv} shows the maximum $\gamma$-photon energy reached in the simulations (second line) and cut-off $\gamma$-photon energies at dN/dE = 284160 MeV$ ^{-1} $ (third line), which is the lower limit in Fig. \ref{fig:angular_distribution_angles}-d.

\begin{table}[h!]
	\caption{\label{table_en_conv} The conversion efficiency of the laser energy into $\gamma$-photon energy, the maximum reached $\gamma$-photon energy and the cut-off $\gamma$-photon energy at dN/dE = 284160 MeV$ ^{-1} $ (the lower limit in Fig. \ref{fig:angular_distribution_angles}-d).} 
	\footnotesize
	\begin{tabular}{@{}l|rrrrrr}
		\br
		Incidence angle & 0.0\ccnd & 30.0\ccnd & 37.5\ccnd & 45.0\ccnd & 52.5\ccnd & 60.0\ccnd \\
		\mr 
		Energy conversion efficiency [\%]  & 3.0 & 5.2 & 6.6 & 8.6 & 10.7 & 8.5 \\		
		Photon maximum energy [MeV] & 370 & 420 & 470 & 450 & 560 & 640 \\
		Photon cut-off energy [MeV]  & 290  & 320  & 340  & 360  & 370  & 370  \\

		\br
	\end{tabular}
	
\end{table}

\section{Conclusion}\label{sec_conclusion}
 In conclusion, we have demonstrated via  3D PIC simulations of the interaction of high-power lasers with solid targets the emission of a collimated $\gamma$-photon beam in the direction parallel to the target surface, for a range of oblique incidence angles. The prior knowledge of the emitted $\gamma$-photons direction is crucial for their detection in future experiments. Moreover, this work will find interest among various applications which typically require a directional $\gamma$-photon beam. The process is ascribed to the formation of interference patterns in the electromagnetic fields by the incident and reflected laser pulse. In this field pattern, the electrons are accelerated to very high energies, while they are temporarily directed parallel to the target surface. In addition to the $\gamma$-photon beam, we observed HHG in the direction parallel to the target surface by the RIME mechanism \cite{Lamaifmmodecheckcelsevcfi2023} in our 3D simulation.

 The collimated patterns prevails in the incidence angle range of 37.5\cc and 52.5\ccnd. The laser energy to $\gamma$-photon energy  conversion efficiency reaches up to 10.7 \%. Although the use of a larger incident angle results in a higher electron energy, the laser energy conversion efficiency into $\gamma$-photons and $\gamma$-photon beam collimation significantly decrease as the angle approaches 60\ccnd. The best $\gamma$-photon beam collimation is obtained for 45\cc incident angle with brilliance approaching $ 10^{23} $ s$^{-1}$mm$^{-2}$mrad$^{-2}$ per 0.1\% bandwidth, which is similar to that of Ref. \cite{Hadjisolomou2022}.

 
  
\ack
The computational time was provided by the supercomputers Sunrise of ELI Beamlines and Karolina of IT4Innovations, supported by the Ministry of Education, Youth and Sports of the Czech Republic through the e-INFRA CZ (ID:90254). The EPOCH code is in part funded by the UK EPSRC grants EP/G054950/1, EP/G056803/1, EP/G055165/1
and EP/M022463/1.

\appendix \label{appendix}
\setcounter{section}{1}
\section*{Appendix A: Dependence on laser polarization, pulse duration, target thickness, preplasma, laser power, and Virtual Reality visualization}

\begin{figure}[ht]
	\begin{center}
		\flushleft
		\includegraphics[width=1\linewidth]{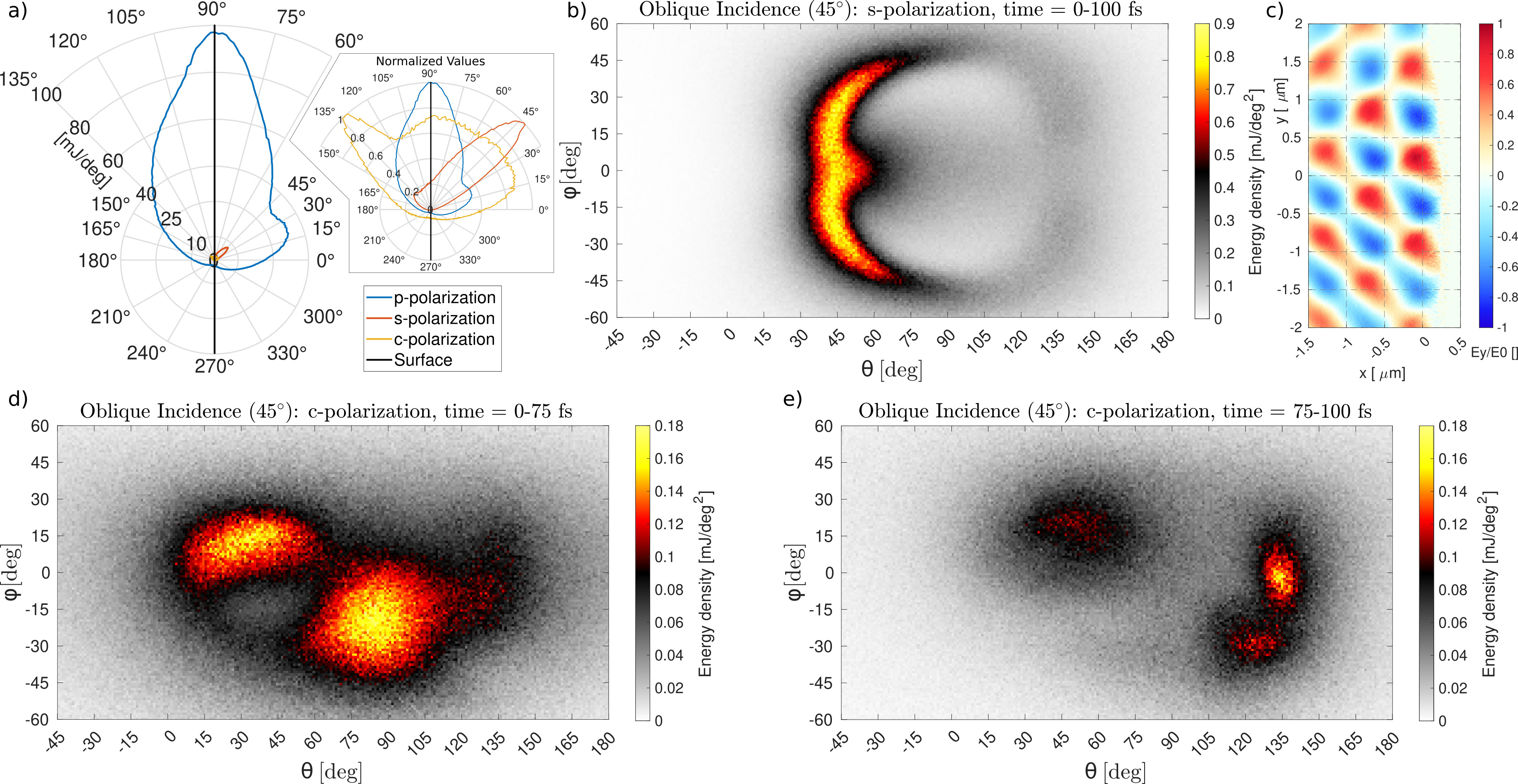}	
		\caption{\label{fig:spherical_polarizations} a) 2D angular distribution of $\gamma$-photon energy in the $ x $-$ y $ plane for the different  polarization cases (the inset shows normalized data). b,d,e) 3D angular energy distribution of the emitted $ \gamma $-photons for b) linear s-polarization for 0-100 fs, d) circular polarization for 0-75 fs , e) circular polarization for 75-100 fs. c) Structures of $ E_y$ at 60 fs for circular polarization at $ x $-$ y $ plane.}
	\end{center}
\end{figure}

Firstly, we show the advantage of using p-polarization (p-pol) over s-polarization (s-pol) and circular polarization (c-pol) in our scheme. Note that for oblique incidence the circularly polarized laser does not interact with the plasma so symmetrically as in the normal incidence case. Fig. \ref{fig:spherical_polarizations}-a shows the angular energy distribution around the incidence plane (eq. \ref{eq:conditionz}) and reveals that the maximum values for p-pol are more than an order of magnitude larger than for s-pol (12.6 times) and c-pol (42.5 times). The beam profiles are normalized in the inset of Fig. \ref{fig:spherical_polarizations}-a. For further investigation we show the 3D angular energy distribution in Figs. \ref{fig:spherical_polarizations}-b,d,e). For the s-pol (Fig. \ref{fig:spherical_polarizations}-b) the energy distribution spreads over 90\cc in the $ \varphi $ direction, as it is the laser polarization plane. In the $ \theta $ direction the distribution is around the incident angle of 45°. However, due to the significant spread in the $ \varphi $ direction the maximum energy density is about an order of magnitude lower than in the p-pol case (see Fig. \ref{fig:spherical_distribution_00x45}-b). For the c-pol, the electric field interference structure in the x-y plane (similar to the p-pol shown in Fig. \ref{fig:field_electron_gamma}(b) but about 1/3 weaker) is shown in Fig. \ref{fig:spherical_polarizations}-c. Therefore, part of the $\gamma$-photons are generated along the target surface ($ \theta=90 $°) in a way similar to that described in Section \ref{sec31} for the p-pol, although more spread in both directions as seen in Fig. \ref{fig:spherical_polarizations}-d. As the mechanism along the target surface is weaker for c-pol, the secondary beam (between  0° and 45°) reaches similar values as the main one (around 90°).  Fig. \ref{fig:spherical_polarizations}-e) shows the $\gamma$-photon energy distribution generated between 75 fs and 100 fs, when the laser pulse is reflected. During this interaction a third beam is generated around the specular angle of 135°.

\begin{figure}[ht]
	\begin{center}
		\flushleft
		\includegraphics[width=1\linewidth]{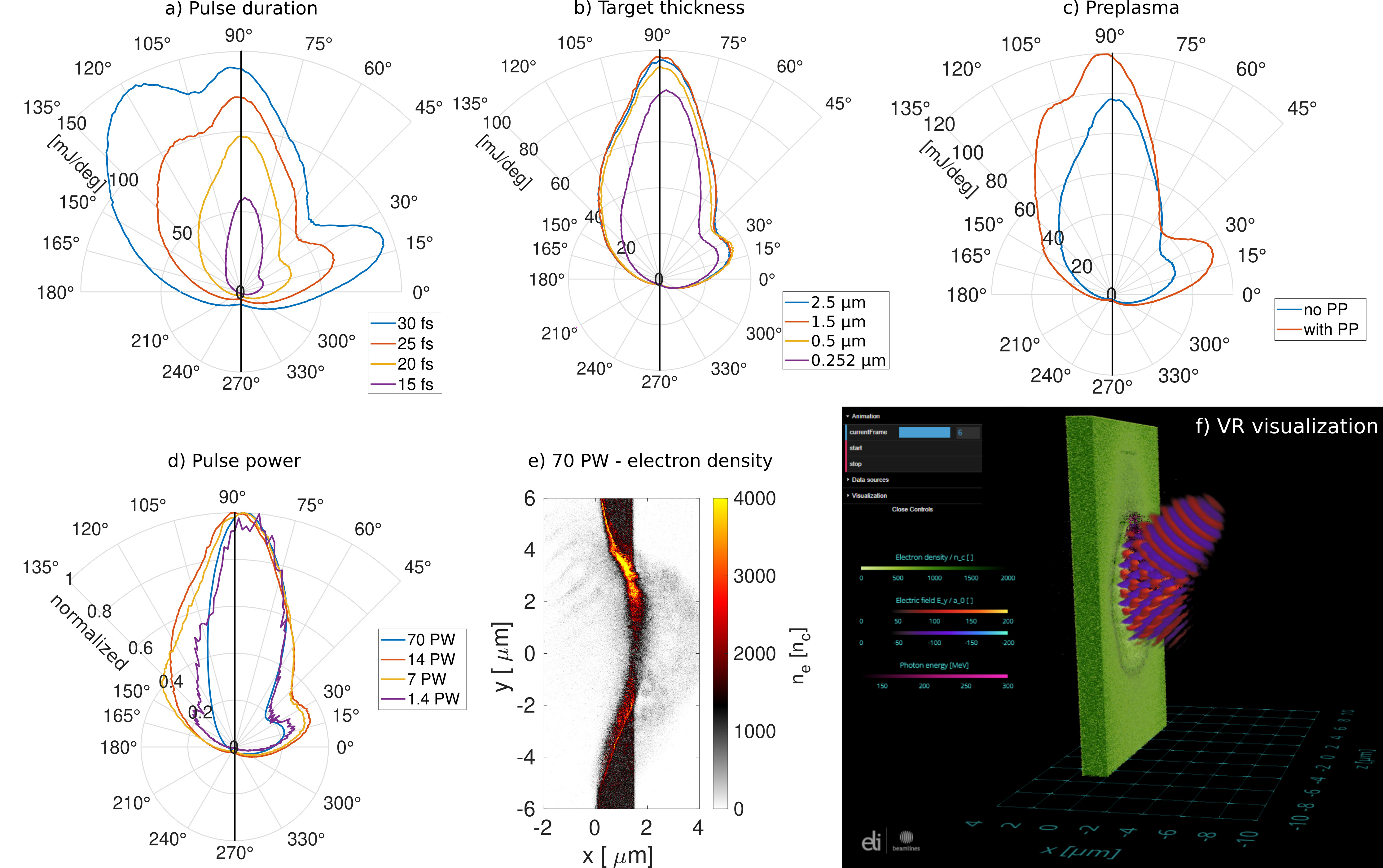}	
		\caption{\label{fig:power_x_thickness} Angular distribution of $\gamma$-photon energy in the $ x $-$ y $ plane. Dependence on a) laser pulse duration, b) target thickness, c) preplasma, d) laser power (normalized data). e) Electron number density at 98.41 fs using a 70 PW laser (colors are saturated). f) VR visualization of the reference case. }
	\end{center}
\end{figure}

Secondly, we compare simulations of different laser pulse durations (Fig. \ref{fig:power_x_thickness}-a), but keeping the same intensity. As expected, longer pulses (larger energy) emit more $ \gamma$-photons. The dominant central lobe along the target surface occurs for all tested pulse durations. For a 15 fs laser pulse, the angular spread is lower than the reference case of 20 fs. On the contrary, for longer pulses, an additional lobe  develops close to the specular angle. This behavior resembles that of increasing the incident angle over 45\cc (\ref{fig:angular_distribution_angles}-e) and is ascribed to target bending (by the rising part of the laser pulse), which changes the actual incidence angle. The target bending is mitigated for lasers with lower intensity.

Thirdly, the effect of the target thickness is presented in Fig. \ref{fig:power_x_thickness}-b. The figure exhibits minimal changes for target thicknesses ranging between 0.5 and 2.5 $\mathrm{\mu m}$. For those thicknesses the iron target is sturdy enough to prevent the 14 PW laser from burning through the target, before the main $\gamma$-photon emission occurs. Decreasing the target thickness to 0.252 $\mathrm{\mu m}$ results in the pulse to penetrate the target, thus limiting the $\gamma$-photon emission.

Fourthly, the effect of the preplasma is presented in Fig. \ref{fig:power_x_thickness}-c, since it was shown that an appropriate preplasma can reduce the divergence angle of the electron beam \cite{Ma2018} and thus influence the $\gamma$-photon generation. We reduced the target thickness to 1 $ \mathrm{\mu m} $ and introduced a 1.75 $ \mathrm{\mu m} $ preplasma in front of it (positions $ x < 0 $) with density ranging from 0.0013 $ n_c $ to the target density. The preplasma  profile is $n_e\exp(-|x|/L) $ with the preplasma scale length $ L = 0.127$ $ \mathrm{\mu m} $. The value of $ L $ is chosen to follow the condition $ L = c/\omega $. This condition is beneficial for generation of high-order harmonics \cite{Dollar2013}. The angular spectra shows a rise in the direction along the target surface by about 23\%.

Lastly in Fig. \ref{fig:power_x_thickness}-d, we compare simulations with different laser power (intensity), while keeping the pulse duration and focal spot the same. The laser power ranges from 1.4 PW to 70 PW  (peak intensity from $ 2\times10^{22}\  \mathrm{W/cm^2} $ to $ 10^{24}\  \mathrm{W/cm^2} $). The values of the figure are normalized. The beam profile of 7 PW case is similar to the 14 PW one, although the  maximum energy density is 4.7 times lower. In the 1.4 PW case the beam is even more collimated. However, the  maximum energy density is reduced by 300 times. The 70 PW case provides the most collimated beam and the  maximum energy density is 23.3 times larger compared to the 14 PW case. At the end of the simulation the laser pulse penetrates the target, as is shown in Fig. \ref{fig:power_x_thickness}-e. Therefore, the $\gamma$-photon yield can be increased with the use of thicker targets.

Additionally, the visualization of the $\gamma$-photon emission temporal evolution of the 45\cc case  can be accessed online \cite{VBL_mm_gamma}  via our Virtual Beamline -- VBL application \cite{VBL_home_page,Danielova2019,Matys2023}. VBL uses a custom-made WebGL \cite{WebGL} app to render  the visualization inside a regular web browser, as shown in Fig. \ref{fig:power_x_thickness}-f. The application also enables support for Virtual Reality (VR) viewing, which was tested with Oculus Rift S and HP Reverb G2 headsets.

\blockcomment{
\section*{Appendix B: Virtual Reality visualization}
The visualization of the $\gamma$-photon emission temporal evolution of the 45\cc case  can be accessed online \cite{VBL_mm_gamma}  via our Virtual Beamline -- VBL application \cite{VBL_home_page,Danielova2019,Matys2023}. VBL uses a custom-made WebGL \cite{WebGL} application  to render  the visualization inside a regular web browser, as shown in Fig \ref{fig:FigVR}. The application also enables support for Virtual Reality (VR) viewing, which was tested with Oculus Rift S and HP Reverb G2 headsets.  
\begin{figure}[ht]
	\begin{center}
		\includegraphics[width=0.81\linewidth]{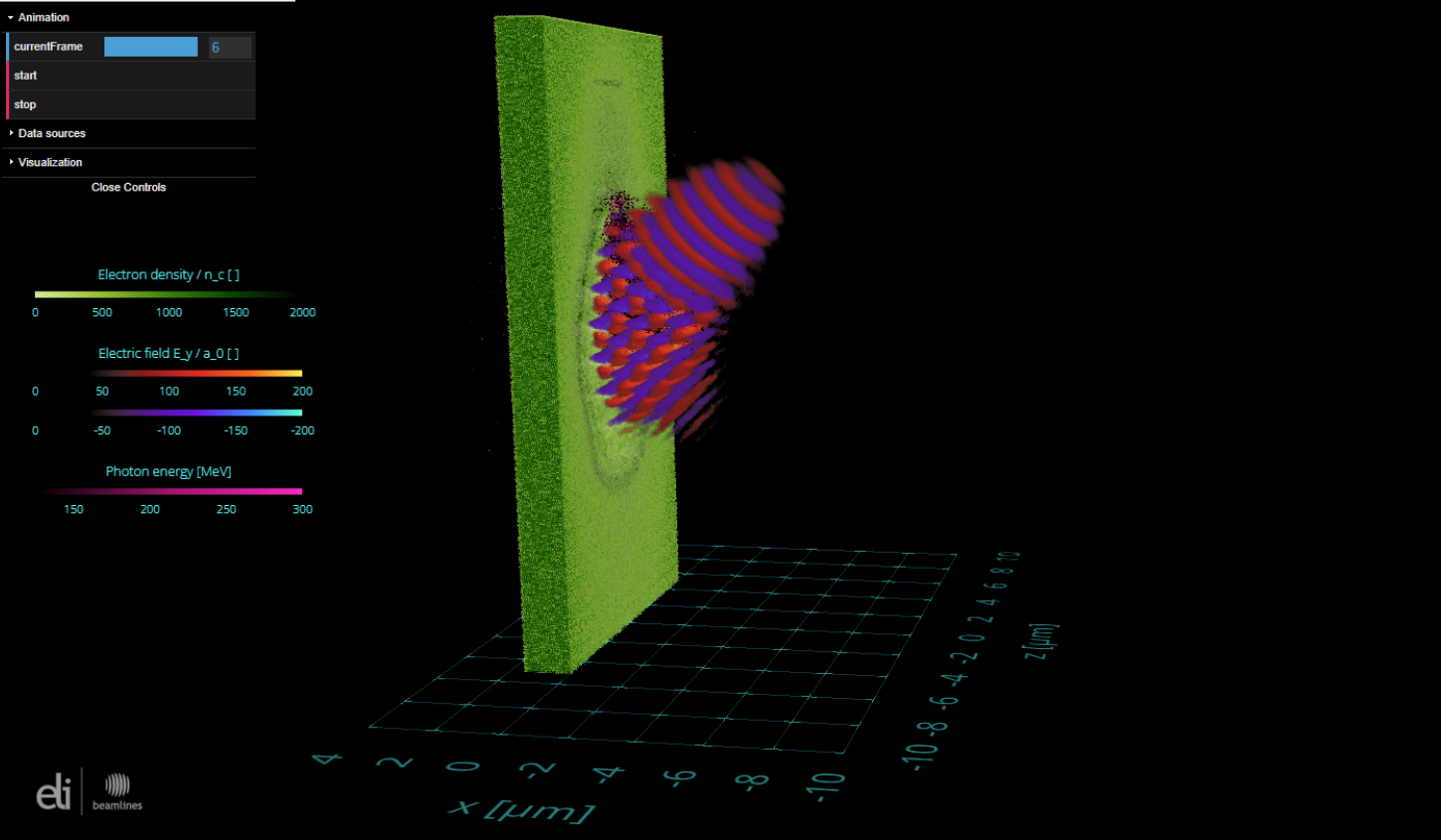}	
		\caption{\label{fig:FigVR} Screenshot of the interactive VBL visualization of the 45\cc incidence case.}
	\end{center}
\end{figure}
}

\section*{References}
\bibliographystyle{iopart-num}
\bibliography{referencesGamma_URL}

\providecommand{\newblock}{}
\begin{thebibliography}{10}
\expandafter\ifx\csname url\endcsname\relax
  \def\url#1{{\tt #1}}\fi
\expandafter\ifx\csname urlprefix\endcsname\relax\def\urlprefix{URL }\fi
\providecommand{\eprint}[2][]{\url{#2}}

\bibitem{Danson2019}
Danson C~N {\em et~al.\/} 2019 {\em High Power Laser Science and Engineering\/}
  {\bf 7} e54 \urlprefix\url{https://doi.org/10.1017/hpl.2019.36}

\bibitem{Tajima1979}
Tajima T and Dawson J~M 1979 {\em Phys. Rev. Lett.\/} {\bf 43}(4) 267--270
  \urlprefix\url{https://doi.org/10.1103/PhysRevLett.43.267}

\bibitem{Esarey2009}
Esarey E, Schroeder C~B and Leemans W~P 2009 {\em Rev. Mod. Phys.\/} {\bf
  81}(3) 1229--1285 \urlprefix\url{https://doi.org/10.1103/RevModPhys.81.1229}

\bibitem{Bulanov2002}
Bulanov S~V, Esirkepov T, Khoroshkov V~S, Kuznetsov A~V and Pegoraro F 2002
  {\em Physics Letters, Section A: General, Atomic and Solid State Physics\/}
  {\bf 299} 240--247
  \urlprefix\url{https://doi.org/10.1016/S0375-9601(02)00521-2}

\bibitem{Daido2012}
Daido H, Nishiuchi M and Pirozhkov A~S 2012 {\em Reports on Progress in
  Physics\/} {\bf 75} 056401
  \urlprefix\url{https://doi.org/10.1088/0034-4885/75/5/056401}

\bibitem{Macchi2013}
Macchi A, Borghesi M and Passoni M 2013 {\em Rev. Mod. Phys.\/} {\bf 85}(2)
  751--793 \urlprefix\url{https://doi.org/10.1103/RevModPhys.85.751}

\bibitem{Bulanov2014}
Bulanov S~V, Wilkens J~J, Esirkepov T~Z, Korn G, Kraft G, Kraft S~D, Molls M
  and Khoroshkov V~S 2014 {\em Physics-Uspekhi\/} {\bf 57} 1149--1179
  \urlprefix\url{https://doi.org/10.3367/ufne.0184.201412a.1265}

\bibitem{Passoni2019}
Passoni M {\em et~al.\/} 2019 {\em Plasma Physics and Controlled Fusion\/} {\bf
  62} 014022 \urlprefix\url{https://doi.org/10.1088/1361-6587/ab56c9}

\bibitem{Teubner2009}
Teubner U and Gibbon P 2009 {\em Reviews of Modern Physics\/} {\bf 81} 445
  \urlprefix\url{https://doi.org/10.1103/RevModPhys.81.445}

\bibitem{Krausz2009}
Krausz F and Ivanov M 2009 {\em Rev. Mod. Phys.\/} {\bf 81}(1) 163--234
  \urlprefix\url{https://doi.org/10.1103/RevModPhys.81.163}

\bibitem{Gonsalves2019}
Gonsalves A~J {\em et~al.\/} 2019 {\em Phys. Rev. Lett.\/} {\bf 122}(8) 084801
  \urlprefix\url{https://doi.org/10.1103/PhysRevLett.122.084801}

\bibitem{Aniculaesei2023}
Aniculaesei C {\em et~al.\/} 2023 {\em Matter and Radiation at Extremes\/} {\bf
  9} 014001 \urlprefix\url{https://doi.org/10.1063/5.0161687}

\bibitem{Ziegler2024}
Ziegler T {\em et~al.\/} 2024 {\em Nature Physics\/} {\bf 20} 1211–1216
  \urlprefix\url{https://doi.org/10.1038/s41567-024-02505-0}

\bibitem{Bulanov2013}
Bulanov S~V, Esirkepov T~Z, Kando M, Pirozhkov A~S and Rosanov N~N 2013 {\em
  Physics-Uspekhi\/} {\bf 56} 429
  \urlprefix\url{https://doi.org/10.3367/UFNe.0183.201305a.0449}

\bibitem{Pirozhkov2017_Biser}
Pirozhkov A~S {\em et~al.\/} 2017 {\em Scientific Reports\/} {\bf 7} 17968
  \urlprefix\url{https://doi.org/10.1038/s41598-017-17498-5}

\bibitem{Lamaifmmodecheckcelsevcfi2023}
Lama\ifmmode~\check{c}\else \v{c}\fi{} M, Mima K, Nejdl J, Chaulagain U and
  Bulanov S~V 2023 {\em Phys. Rev. Lett.\/} {\bf 131}(20) 205001
  \urlprefix\url{https://doi.org/10.1103/PhysRevLett.131.205001}

\bibitem{Mourou2006}
Mourou G~A, Tajima T and Bulanov S~V 2006 {\em Reviews of Modern Physics\/}
  {\bf 78} 309--371 \urlprefix\url{https://doi.org/10.1103/RevModPhys.78.309}

\bibitem{Marklund_Shukla_2006}
Marklund M and Shukla P~K 2006 {\em Rev. Mod. Phys.\/} {\bf 78}(2) 591--640
  \urlprefix\url{https://doi.org/10.1103/RevModPhys.78.591}

\bibitem{DiPiazza_Keitel2012}
Di~Piazza A, M\"uller C, Hatsagortsyan K~Z and Keitel C~H 2012 {\em Rev. Mod.
  Phys.\/} {\bf 84}(3) 1177--1228
  \urlprefix\url{https://doi.org/10.1103/RevModPhys.84.1177}

\bibitem{Gonoskov2022}
Gonoskov A, Blackburn T~G, Marklund M and Bulanov S~S 2022 {\em Rev. Mod.
  Phys.\/} {\bf 94}(4) 045001
  \urlprefix\url{https://doi.org/10.1103/RevModPhys.94.045001}

\bibitem{Burke1997}
Burke D~L {\em et~al.\/} 1997 {\em Phys. Rev. Lett.\/} {\bf 79}(9) 1626--1629
  \urlprefix\url{https://doi.org/10.1103/PhysRevLett.79.1626}

\bibitem{Abramowicz2021}
Abramowicz H {\em et~al.\/} 2021 {\em The European Physical Journal Special
  Topics\/} {\bf 230} 2445–2560
  \urlprefix\url{https://doi.org/10.1140/epjs/s11734-021-00249-z}

\bibitem{Ridgers2012}
Ridgers C~P, Brady C~S, Duclous R, Kirk J~G, Bennett K, Arber T~D, Robinson
  A~P~L and Bell A~R 2012 {\em Phys. Rev. Lett.\/} {\bf 108}(16) 165006
  \urlprefix\url{https://doi.org/10.1103/PhysRevLett.108.165006}

\bibitem{Nakamura2012}
Nakamura T, Koga J~K, Esirkepov T~Z, Kando M, Korn G and Bulanov S~V 2012 {\em
  Phys. Rev. Lett.\/} {\bf 108}(19) 195001
  \urlprefix\url{https://doi.org/10.1103/PhysRevLett.108.195001}

\bibitem{Lezhnin2018}
Lezhnin K~V, Sasorov P~V, Korn G and Bulanov S~V 2018 {\em Physics of
  Plasmas\/} {\bf 25} 123105 \urlprefix\url{https://doi.org/10.1063/1.5062849}

\bibitem{Hadjisolomou2023_review}
Hadjisolomou P, Jeong T~M, Kolenaty D, Macleod A~J, Olšovcová V, Versaci R,
  Ridgers C~P and Bulanov S~V 2023 {\em Physics of Plasmas\/} {\bf 30} 093103
  \urlprefix\url{https://doi.org/10.1063/5.0158264}

\bibitem{Pirozhkov_2024_arxiv}
Pirozhkov A~S {\em et~al.\/} 2024 {\em arXiv:2410.06537\/} (\textit{Preprint}
  \eprint{https://doi.org/10.48550/ARXIV.2410.06537})

\bibitem{Eliasson2013}
Eliasson B and Liu C~S 2013 {\em Journal of Plasma Physics\/} {\bf 79}
  995–998 \urlprefix\url{https://doi.org/10.1017/S0022377813000779}

\bibitem{Albert_2016}
Albert F and Thomas A~G~R 2016 {\em Plasma Physics and Controlled Fusion\/}
  {\bf 58} 103001
  \urlprefix\url{https://doi.org/10.1088/0741-3335/58/10/103001}

\bibitem{Ledingham2000_Photonuclear_Phys}
Ledingham K~W~D {\em et~al.\/} 2000 {\em Phys. Rev. Lett.\/} {\bf 84}(5)
  899--902 \urlprefix\url{https://doi.org/10.1103/PhysRevLett.84.899}

\bibitem{Nedorezov2004}
Nedorezov V~G, Turinge A~A and Shatunov Y~M 2004 {\em Physics-Uspekhi\/} {\bf
  47} 341 \urlprefix\url{https://doi.org/10.1070/PU2004v047n04ABEH001743}

\bibitem{Kolenaty2022}
Kolenat\'y D, Hadjisolomou P, Versaci R, Jeong T~M, Valenta P,
  Ol\ifmmode~\check{s}\else \v{s}\fi{}ovcov\'a V and Bulanov S~V 2022 {\em
  Phys. Rev. Res.\/} {\bf 4}(2) 023124
  \urlprefix\url{https://doi.org/10.1103/PhysRevResearch.4.023124}

\bibitem{Pomerantz2014_neutron}
Pomerantz I {\em et~al.\/} 2014 {\em Phys. Rev. Lett.\/} {\bf 113}(18) 184801
  \urlprefix\url{https://doi.org/10.1103/PhysRevLett.113.184801}

\bibitem{Cowan2000_fission}
Cowan T~E {\em et~al.\/} 2000 {\em Phys. Rev. Lett.\/} {\bf 84}(5) 903--906
  \urlprefix\url{https://doi.org/10.1103/PhysRevLett.84.903}

\bibitem{Schwoerer2003}
Schwoerer H, Ewald F, Sauerbrey R, Galy J, Magill J, Rondinella V, Schenkel R
  and Butz T 2003 {\em Europhysics Letters\/} {\bf 61} 47
  \urlprefix\url{https://doi.org/10.1209/epl/i2003-00243-1}

\bibitem{Weeks1997}
Weeks K~J, Litvinenko V~N and Madey J~M~J 1997 {\em Medical Physics\/} {\bf 24}
  417--423 \urlprefix\url{https://doi.org/10.1118/1.597903}

\bibitem{Antonelli2017}
Antonelli L {\em et~al.\/} 2017 {\em Phys. Rev. E\/} {\bf 95}(6) 063205
  \urlprefix\url{https://doi.org/10.1103/PhysRevE.95.063205}

\bibitem{Ehlotzky2009}
Ehlotzky F, Krajewska K and Kamiński J~Z 2009 {\em Reports on Progress in
  Physics\/} {\bf 72} 046401
  \urlprefix\url{https://doi.org/10.1088/0034-4885/72/4/046401}

\bibitem{Maslarova2023}
Maslarova D, Martinez B and Vranic M 2023 {\em Physics of Plasmas\/} {\bf 30}
  093107 \urlprefix\url{https://doi.org/10.1063/5.0160121}

\bibitem{Martinez2023}
Martinez B, Barbosa B and Vranic M 2023 {\em Phys. Rev. Accel. Beams\/} {\bf
  26}(1) 011301
  \urlprefix\url{https://doi.org/10.1103/PhysRevAccelBeams.26.011301}

\bibitem{Ujeniuc_Suvaila_2024}
Ujeniuc S and Suvaila R 2024 {\em EPJ Quantum Technology\/} {\bf 11} 39
  \urlprefix\url{https://doi.org/10.1140/epjqt/s40507-024-00240-2}

\bibitem{Bulanov_Laboratory_Astrophysics_2015}
Bulanov S~V, Esirkepov T~Z, Kando M, Koga J, Kondo K and Korn G 2015 {\em
  Plasma Physics Reports\/} {\bf 41} 1–51
  \urlprefix\url{https://doi.org/10.1134/s1063780x15010018}

\bibitem{Rees1992_fireball}
Rees M~J and Mészáros P 1992 {\em Monthly Notices of the Royal Astronomical
  Society\/} {\bf 258} 41P--43P
  \urlprefix\url{https://doi.org/10.1093/mnras/258.1.41P}

\bibitem{Philippov2018_Pulsar}
Philippov A~A and Spitkovsky A 2018 {\em The Astrophysical Journal\/} {\bf 855}
  94 \urlprefix\url{https://doi.org/10.3847/1538-4357/aaabbc}

\bibitem{Aharonian_2021_Astrophysics}
Aharonian F {\em et~al.\/} 2021 {\em Phys. Rev. Lett.\/} {\bf 126}(24) 241103
  \urlprefix\url{https://doi.org/10.1103/PhysRevLett.126.241103}

\bibitem{Zhidkov2002}
Zhidkov A, Koga J, Sasaki A and Uesaka M 2002 {\em Phys. Rev. Lett.\/} {\bf
  88}(18) 185002 \urlprefix\url{https://doi.org/10.1103/PhysRevLett.88.185002}

\bibitem{Koga2005}
Koga J, Esirkepov T~Z and Bulanov S~V 2005 {\em Physics of Plasmas\/} {\bf 12}
  093106 \urlprefix\url{https://doi.org/10.1063/1.2013067}

\bibitem{Gu_Klimo_Bulanov_Weber_2018}
Gu Y~J, Klimo O, Bulanov S~V and Weber S 2018 {\em Communications Physics\/}
  {\bf 1} 93 \urlprefix\url{https://doi.org/10.1038/s42005-018-0095-3}

\bibitem{Bell2008}
Bell A~R and Kirk J~G 2008 {\em Phys. Rev. Lett.\/} {\bf 101}(20) 200403
  \urlprefix\url{https://doi.org/10.1103/PhysRevLett.101.200403}

\bibitem{Kirk2009}
Kirk J~G, Bell A~R and Arka I 2009 {\em Plasma Physics and Controlled Fusion\/}
  {\bf 51} 085008 \urlprefix\url{https://doi.org/10.1088/0741-3335/51/8/085008}

\bibitem{Luo2015}
Luo W, Zhu Y~B, Zhuo H~B, Ma Y~Y, Song Y~M, Zhu Z~C, Wang X~D, Li X~H, Turcu
  I~C~E and Chen M 2015 {\em Physics of Plasmas\/} {\bf 22} 063112
  \urlprefix\url{https://doi.org/10.1063/1.4923265}

\bibitem{Grismayer2016}
Grismayer T, Vranic M, Martins J~L, Fonseca R~A and Silva L~O 2016 {\em Physics
  of Plasmas\/} {\bf 23} 056706
  \urlprefix\url{https://doi.org/10.1063/1.4950841}

\bibitem{Hadjisolomou2025_wire}
Hadjisolomou P, Jeong T~M, Valenta P, Macleod A~J, Shaisultanov R, Ridgers C~P
  and Bulanov S~V 2025 {\em Phys. Rev. E\/} {\bf 111}(2) 025201
  \urlprefix\url{https://doi.org/10.1103/PhysRevE.111.025201}

\bibitem{Vranic2016}
Vranic M, Grismayer T, Fonseca R~A and Silva L~O 2016 {\em Plasma Physics and
  Controlled Fusion\/} {\bf 59} 014040
  \urlprefix\url{https://doi.org/10.1088/0741-3335/59/1/014040}

\bibitem{Gong2017}
Gong Z, Hu R~H, Shou Y~R, Qiao B, Chen C~E, He X~T, Bulanov S~S, Esirkepov T~Z,
  Bulanov S~V and Yan X~Q 2017 {\em Phys. Rev. E\/} {\bf 95}(1) 013210
  \urlprefix\url{https://doi.org/10.1103/PhysRevE.95.013210}

\bibitem{Nerush2014}
Nerush E~N, Kostyukov I~Y, Ji L and Pukhov A 2014 {\em Physics of Plasmas\/}
  {\bf 21} 013109 \urlprefix\url{https://doi.org/10.1063/1.4863423}

\bibitem{Stark2016}
Stark D~J, Toncian T and Arefiev A~V 2016 {\em Phys. Rev. Lett.\/} {\bf
  116}(18) 185003
  \urlprefix\url{https://doi.org/10.1103/PhysRevLett.116.185003}

\bibitem{Wang2020_Double_Lobe}
Wang T, Ribeyre X, Gong Z, Jansen O, d'Humi\`eres E, Stutman D, Toncian T and
  Arefiev A 2020 {\em Phys. Rev. Appl.\/} {\bf 13}(5) 054024
  \urlprefix\url{https://doi.org/10.1103/PhysRevApplied.13.054024}

\bibitem{Wang2020}
Wang X~B, Hu G~Y, Zhang Z~M, Gu Y~Q, Zhao B, Zuo Y and Zheng J 2020 {\em High
  Power Laser Science and Engineering\/} {\bf 8} e34
  \urlprefix\url{https://doi.org/10.1017/hpl.2020.30}

\bibitem{Vyskocil2020}
Vyskočil J, Gelfer E and Klimo O 2020 {\em Plasma Physics and Controlled
  Fusion\/} {\bf 62} 064002
  \urlprefix\url{https://doi.org/10.1088/1361-6587/ab83cb}

\bibitem{Hadjisolomou2021}
Hadjisolomou P, Jeong T~M, Valenta P, Korn G and Bulanov S~V 2021 {\em Phys.
  Rev. E\/} {\bf 104}(1) 015203
  \urlprefix\url{https://doi.org/10.1103/PhysRevE.104.015203}

\bibitem{Hadjisolomou_2022_longer}
Hadjisolomou P, Jeong T, Valenta P, Kolenaty D, Versaci R, Olšovcová V,
  Ridgers C and Bulanov S 2022 {\em Journal of Plasma Physics\/} {\bf 88}
  905880104 \urlprefix\url{https://doi.org/10.1017/S0022377821001318}

\bibitem{Hadjisolomou2022}
Hadjisolomou P, Jeong T~M and Bulanov S~V 2022 {\em Scientific Reports\/} {\bf
  12} 17143 \urlprefix\url{https://doi.org/10.1038/s41598-022-21352-8}

\bibitem{Galbiati2023}
Galbiati M, Formenti A, Grech M and Passoni M 2023 {\em Frontiers in Physics\/}
  {\bf 11} \urlprefix\url{https://doi.org/10.3389/fphy.2023.1117543}

\bibitem{Formenti2024}
Formenti A, Galbiati M and Passoni M 2024 {\em Phys. Rev. E\/} {\bf 109}(3)
  035206 \urlprefix\url{https://doi.org/10.1103/PhysRevE.109.035206}

\bibitem{Kleij2024}
Kleij P~S, Marini S, Caetano~de Sousa M, Grech M, Riconda C and Raynaud M 2024
  {\em Physics of Plasmas\/} {\bf 31} 072111
  \urlprefix\url{https://doi.org/10.1063/5.0209316}

\bibitem{Goodman_McKenna_2023}
Goodman J, King M, Dolier E~J, Wilson R, Gray R~J and McKenna P 2023 {\em High
  Power Laser Science and Engineering\/} {\bf 11} e34
  \urlprefix\url{https://doi.org/10.1017/hpl.2023.11}

\bibitem{Serebryakov2016}
Serebryakov D and Nerush E 2016 {\em Quantum Electronics\/} {\bf 46} 299
  \urlprefix\url{https://doi.org/10.1070/QEL16051}

\bibitem{Cantono_Macchi_2018}
Cantono G, Fedeli L, Sgattoni A, Denoeud A, Chopineau L, R\'eau F, Ceccotti T
  and Macchi A 2018 {\em Phys. Rev. Lett.\/} {\bf 120}(26) 264803
  \urlprefix\url{https://doi.org/10.1103/PhysRevLett.120.264803}

\bibitem{Shen_Pukhov_Qiao_2024}
Shen X, Pukhov A and Qiao B 2024 {\em Communications Physics\/} {\bf 7} 84
  \urlprefix\url{https://doi.org/10.1038/s42005-024-01575-z}

\bibitem{Sarma_Macchi_2022}
Sarma J, McIlvenny A, Das N, Borghesi M and Macchi A 2022 {\em New Journal of
  Physics\/} {\bf 24} 073023
  \urlprefix\url{https://doi.org/10.1088/1367-2630/ac7d6e}

\bibitem{Chen2006}
Chen M, Shenga Z~M, Zheng J, Ma Y~Y, Bari M~A, Li Y~T and Zhang J 2006 {\em
  Opt. Express\/} {\bf 14} 3093--3098
  \urlprefix\url{https://doi.org/10.1364/OE.14.003093}

\bibitem{Serebryakov_2017_near_surface_el}
Serebryakov D~A, Nerush E~N and Kostyukov I~Y 2017 {\em Physics of Plasmas\/}
  {\bf 24} 123115 \urlprefix\url{https://doi.org/10.1063/1.5002671}

\bibitem{Arber2015}
Arber T~D {\em et~al.\/} 2015 {\em Plasma Physics and Controlled Fusion\/} {\bf
  57} 113001 \urlprefix\url{https://doi.org/10.1088/0741-3335/57/11/113001}

\bibitem{KOCH1959}
Koch H~W and Motz J~W 1959 {\em Rev. Mod. Phys.\/} {\bf 31}(4) 920--955
  \urlprefix\url{https://doi.org/10.1103/RevModPhys.31.920}

\bibitem{Vyskocil2018_Brems}
Vyskočil J, Klimo O and Weber S 2018 {\em Plasma Physics and Controlled
  Fusion\/} {\bf 60} 054013
  \urlprefix\url{https://doi.org/10.1088/1361-6587/aab4c3}

\bibitem{VBL_mm_gamma}
Collimated $\gamma$-flash emission along the target surface
  \url{https://vbl.eli-beams.eu/mm-gamma/}
  \urlprefix\url{https://vbl.eli-beams.eu/mm-gamma/}

\bibitem{VBL_home_page}
Virtual beamline \url{https://vbl.eli-beams.eu/}
  \urlprefix\url{https://vbl.eli-beams.eu/}

\bibitem{Danielova2019}
Danielova M, Janecka P, Grosz J and Holy A 2019 {Interactive 3D Visualizations
  of Laser Plasma Experiments on the Web and in VR} {\em EuroVis 2019 -
  Posters\/} (The Eurographics Association)
  \urlprefix\url{https://doi.org/10.2312/eurp.20191145}

\bibitem{Matys2023}
Matys M, Psikal J, Nishihara K, Klimo O, Jirka M, Valenta P and Bulanov S~V
  2023 {\em Photonics\/} {\bf 10} 61
  \urlprefix\url{https://doi.org/10.3390/photonics10010061}

\bibitem{Baier1998}
Baier V~N, Katkov V~M and Strakhovenko V~M 1998 {\em Electromagnetic Processes
  at High Energies in Oriented Single Crystals\/} (Singapore: World Scientific)
  \urlprefix\url{https://doi.org/10.1142/2216}

\bibitem{Bulanov1994_ROM}
Bulanov S~V, Naumova N~M and Pegoraro F 1994 {\em Physics of Plasmas\/} {\bf 1}
  745--757 \urlprefix\url{https://doi.org/10.1063/1.870766}

\bibitem{Grassi2017}
Grassi A, Grech M, Amiranoff F, Macchi A and Riconda C 2017 {\em Phys. Rev.
  E\/} {\bf 96}(3) 033204
  \urlprefix\url{https://doi.org/10.1103/PhysRevE.96.033204}

\bibitem{Thevenet_2015}
Thévenet M, Leblanc A, Kahaly S, Vincenti H, Vernier A, Quéré F and Faure J
  2015 {\em Nature Physics\/} {\bf 12} 355–360
  \urlprefix\url{https://doi.org/10.1038/nphys3597}

\bibitem{Ma2018}
Ma Y, Zhao J, Li Y, Li D, Chen L, Liu J, Dann S~J~D, Ma Y, Yang X, Ge Z, Sheng
  Z and Zhang J 2018 {\em Proceedings of the National Academy of Sciences\/}
  {\bf 115} 6980–6985 \urlprefix\url{https://doi.org/10.1073/pnas.1800668115}

\bibitem{Dollar2013}
Dollar F, Cummings P, Chvykov V, Willingale L, Vargas M, Yanovsky V, Zulick C,
  Maksimchuk A, Thomas A~G~R and Krushelnick K 2013 {\em Phys. Rev. Lett.\/}
  {\bf 110}(17) 175002
  \urlprefix\url{https://doi.org/10.1103/PhysRevLett.110.175002}

\bibitem{WebGL}
Parisi T 2012; {\em WebGL: Up and Running\/} 1st ed (Sebastopol, CA, USA:
  O'Reilly Media, Inc.)

\end{thebibliography}
\end{document}